\documentclass[aps,prx,reprint,superscriptaddress,noeprint,longbibliography,nofootinbib]{revtex4-2}

\usepackage{enumitem}
\usepackage[normalem]{ulem}
\usepackage{amsmath}

\usepackage{mathrsfs}
\usepackage{amsbsy}
\usepackage{hhline}
\usepackage{multirow}
\usepackage{amssymb} 
\usepackage{graphicx}
\usepackage{float}
\usepackage{upgreek}
\usepackage[caption=false]{subfig}
\usepackage{hyperref}
\hypersetup{
	colorlinks=true,       
	linkcolor=magenta,          
	citecolor=blue,       
	filecolor=magenta,      
	urlcolor=blue          
}

\usepackage[capitalise]{cleveref}

\usepackage[dvipsnames]{xcolor}

\newcommand{\bxi}{{\boldsymbol \xi}}

\newcommand{\im}{{\rm Im\,}}
\newcommand{\re}{{\rm Re\,}}

\newcommand{\tr}{{\rm Tr}}

\newcommand{\llangle}{{\langle\!\langle}}
\newcommand{\rrangle}{{\rangle\!\rangle}}

\usepackage{blkarray}% http://ctan.org/pkg/blkarray
\newcommand{\bX}{{\mathbf X}}

\newcommand{\bsigma}{{\boldsymbol \sigma}}

\begin{document}
	
\title{
Hybrid Quantum-Classical Stochastic Approach to Spin-Boson Models}	

\author{Naushad A. Kamar}

\author{Mohammad Maghrebi}
    \email[Corresponding author: ]{maghrebi@msu.edu}
\affiliation{Department of Physics and Astronomy, Michigan State University, East Lansing, Michigan 48824 USA}

\date{\today}

\begin{abstract}
    Interacting spin-boson models encompass a large class of physical systems, spanning models with a single spin interacting with a bosonic bath---a paradigm of quantum impurity problems---to models with many spins interacting with a cavity mode---a paradigm of quantum optics.  Such models have emerged in various quantum simulation platforms which are further subject to noise and lossy dynamics. As generic many-body systems, dynamics of spin-boson models constitutes a challenging problem. 
    In this paper, we present an exact hybrid quantum-classical stochastic approach to different spin-boson models which are typically treated using distinct techniques. 
    In this approach, the solution of a classical stochastic equation (mimicking the bosonic modes) is input into a quantum stochastic equation for the spins. Furthermore, the spins are effectively decoupled for each stochastic realization, but this comes at the expense of sampling over unphysical states.
    Remarkably, the dynamics remains Markovian in our approach even in the strong coupling regime. Moreover, we utilize Markovian dissipation to make \textit{causality} manifest, thus ensuring hermiticity (though not positivity) of the density matrix for each realization. Finally, in contrast with many existing methods, we place no restriction on the initial state, and further argue that an intrinsic nonlinearity of the bosonic modes can be tackled within this framework. We benchmark and showcase the utility of our approach in several examples, specifically in cases where an exact numerical calculation is far from reach. 
\end{abstract}


\maketitle

\section{Introduction}

Spin-boson models where one or many spins interact with bosonic modes encompass a large class of physical models. On one hand, the paradigmatic spin-boson model describing  a single two-level system coupled to an (infinite) bosonic bath \cite{Leggett_spin_boson_model,weiss2012quantum} defines a paradigm of quantum impurity problems, with applications to physical, chemical and even biological problems \cite{XU199491,garg1985effect,weiss2012quantum,Leggett_spin_boson_model,Bulla_2008}.
On the other hand, the Dicke model describing many spins coupled to a cavity mode defines a paradigm of quantum optics and gives rise to a superradiant phase transition at strong coupling \cite{Lieb73,Hepp73}. Such models are particularly relevant in the setting of quantum simulation platforms: the paradigmatic spin-boson model has  been recently realized in superconducting qubits \cite{Diaz_Ultra_Strong_Spin_Boson_model,yoshihara2017superconducting}, while the Dicke model
has been implemented in multiple platforms \cite{Baumann2010,Klinder_2015,Zhiqiang:17,Safavi-Naini_2018,Muniz_2020}; quantum Rabi models, describing a single spin coupled to a boson, are also demonstrated recently in trapped ions
\cite{Duan_2018,Duan_2021,Duan_2022}. Furthermore, generalized spin-boson models involving many interacting spins and bosons can be implemented in various platforms ranging from trapped ions \cite{Sorensen_1999,Porras_2004,Porras_2008,Kim_2009,britton2012engineered}, to cavities via cold atoms \cite{Leroux_2010,Ritsch_2013,Vaidya_2018}, to superconducting qubits \cite{niemczyk2010circuit,LEHUR2016808}, and optomechanics \cite{Aspelmeyer_2014}. 
Quantum simulation platforms are further subject to incoherent dynamics, for example, due to 
the noise in lasers or the cavity loss.
A full description then accounts for dissipative dynamics, and in many settings (specifically, those involving cavities) takes the form of a quantum Markovian master equation.
The resulting dynamics gives rise to driven-dissipative systems
defined by the competition between a coherent drive and
incoherent loss 
\cite{Diehl2008,verstraete2009quantum}. Driven-dissipative dynamics of interacting spin-boson models emerges in various settings such as trapped ions \cite{Schindler13,Bohnet16,Monroe_2021}, Rydberg gases \cite{Carr13,Malossi2014}, circuit QED  \cite{Houck2012,Fitzpatrick2017}, and cavity QED platforms \cite{Baden_2014,Landini_2018,Guo_2021}.

Dynamics of many-body systems out of equilibrium constitutes a notoriously challenging problem, as the size of the Hilbert space is exponentially large, and traditional Monte Carlo methods suffer from the dynamical sign problem. Spin-boson models are further complicated due to the unbounded local Hilbert space of bosons \cite{Steven_White_Optimal_Basis,kamar_splitting_of_Hilbert_Space}.
Moreover, 
memory effects become important at strong coupling, 
leading to non-Markovian dynamics of spins. 
Distinct techniques have been devised for different types of spin-boson models: for the paradigmatic spin-boson model, these techniques range from perturbative analytical methods such as NIBA \cite{weiss2012quantum,Leggett_spin_boson_model}, to various stochastic methods \cite{Strunz_1999,Stockburger1998,Peter_Stochastic_Schrodinger_Equation,deVega_2017}, Monte Carlo simulations \cite{Egger1994,MAKAROV1994482}, matrix product state (MPS) based methods \cite{Plenio_Spin_Boson2010,Wall_2016,Keeling_MPDO}, and more recently non-Gaussian variational ansatze \cite{Knorzer_2022,bond2023fast}. On the other hand, many-body spin systems coupled to a cavity mode have been treated 
using mean field theory and cumulant expansion \cite{Kirton_2017,Reiter_2020}, (discrete)~truncated Wigner approximation \cite{Pineiro_2017,Kelly_2021}, (Keldysh) field theory methods \cite{DallaTorre13,Sieberer2016,Paz_2021_exact}, 
quantum trajectories \cite{Daley_2014}, and tensor networks \cite{Weimer_Open_Quantum_System_Method_Review,Mahmoodian_2020}. Finally, exact solutions via Bethe ansatz \cite{HUR20082208} or alternative methods \cite{McDonald_2022} are available in special cases of these models. 
Simple descriptions via a (convolutionless, or Redfield, e.g.) master equation involving only the spins can be obtained in certain regimes \cite{breuer2002theory,Damanet_2019};
however, they are generally unavailable specifically in the strong coupling regime, and, even when they are, the resulting model remains challenging for a many-body spin system. 

In this work, we consider a generalized spin-boson model and develop a hybrid stochastic quantum-classical approach to the evolution of the spins. 
In this approach, we first solve a classical stochastic equation (mimicking the bosonic modes) whose solution is then fed into a quantum stochastic equation for the spins. 
Furthermore, the spins are effectively decoupled for each stochastic realization, but  this comes at the expense of sampling over unphysical (e.g., non-positive) states.
Our work provides a uniform approach to different spin-boson models, and is distinct from previous stochastic approaches in several characteristic ways: i) our hybrid quantum-classical approach remains Markovian even in the strong coupling regime; ii) causality is made manifest (thanks to Markovian loss), preserving the hermiticity of the density matrix for each realization; iii) no restriction is placed on the initial state. Interestingly, the second feature indicates that Markovian dissipation can be used as a resource for numerical simulation. 
Furthermore, we argue that our approach can tackle intrinsic nonlinearities of the bosonic modes. 
We benchmark our method and showcase its utility in several examples where an exact numerical computation is far from reach. 

The structure of this paper is as follows: In \cref{sec:Model}, we introduce the generalized spin-boson model, and further discuss a first stochastic approach to decoupling the spins while pointing to its limitations. In \cref{sec:main_results}, we summarize the main results of the paper. In \cref{sec:N=M=1}, we present our main approach in application to a model comprising a single spin coupled to a cavity mode. In \cref{sec:N>1_M=1}, we generalize this treatment to many spins. In \cref{sec:N=1_M>1}, we consider a single spin coupled to many bosonic modes, and finally in \cref{sec:N>1_M>1}, we consider the generalized model with many interacting spins and bosons. We summarize and outline future directions in \cref{sec:outlook}. The technical derivation of the main results is provided in \cref{sec:derivation_rho,sec:derivation_rho_xi,sec:FV,sec:checkC}.

\section{Model}\label{sec:Model}
We consider a generalized model where spins  are coupled to one or many bosonic modes, described by the Hamiltonian 
\begin{equation}\label{eq:hamiltonian}
    H_{N,M}= \frac{\Delta}{2} \sum_{i=1}^N \sigma^z_i +\sum_{\alpha=1}^M\omega_\alpha a^\dagger_\alpha a_\alpha +  \frac{1}{\sqrt{N}}\sum_{\alpha i}\frac{g_{\alpha i}}{2} \sigma^x_i(a_\alpha+a_\alpha^\dagger)
\end{equation}
where the spin on site $i$ is coupled linearly to the bosonic operators $a_\alpha$ via the coupling $g_{\alpha i}$. 
The overall normalization factor in front of the last term in the Hamiltonian is included for convenience and renders the Hamiltonian extensive in the number of spins. We shall denote the three terms on the rhs as $H_S$, $H_B$, and $H_{SB}$, respectively. 
A prototypical example is a cavity QED system where atoms are placed inside a multimode cavity \cite{Vaidya_2018}. 
Additionally, we assume that the system is lossy; in the example of cavity QED, this could be the cavity loss.
Under the Born-Markov approximation, the dynamics is governed by a quantum Markovian master equation given by \cite{gorini1976completely,lindblad1976generators}
\begin{align}\label{eq:liouvillian}
\begin{split}
        \frac{d\rho}{dt} = & {\cal L}(\rho) =-i[H_{N,M},\rho] + \\
     &+\sum_\alpha 2L_\alpha\rho L_\alpha^\dagger-\rho L_\alpha^\dagger L_\alpha-L_\alpha^\dagger L_\alpha \rho 
\end{split}
\end{align}
Here, $\rho$ is the density matrix of the full system and  $\cal L$ defines the Liouvillian comprising both the Hamiltonian and the dissipative dynamics.
The operators
$L_\alpha$ are the so-called Lindblad operators, and characterize dissipation.
We assume that the bosonic modes are subject to loss at the rate $\kappa_\alpha$ with the corresponding Lindblad operator $L_\alpha = \sqrt{\kappa_\alpha} a_\alpha$. Additionally, spins could be subject to loss, for example, via atomic spontaneous emission. 
We emphasize that $H_{N,M}$ in the above equation should be interpreted as the Hamiltonian in the rotating frame, and the driven nature of the model is thus disguised in the rotating frame. 
We are particularly interested in the reduced density matrix of the spins $\rho_S = \tr_B(\rho)$. In the absence of the coupling to the bosonic modes, the spin dynamics is generated by  the spin-only part of the full Liouvillian  denoted by ${\cal L}_S$. Since spins do not interact directly, the latter is a sum of local terms, 
${\cal L}_S = \sum_i {\cal L}_i$.

\subsection*{Decoupling spins: A first approach}

Many studies of the spin-boson models assume that the initial state  is factorized (i.e., spins and bosons are initially uncorrelated) and furthermore the bosonic modes are initially thermal (hence, Gaussian). 
With this assumption, one can exactly trace out the bosonic modes thanks to the path-integral formulation by Feynman and Vernon \cite{Feynman-Vernon,weiss2012quantum,Leggett_spin_boson_model}. This approach makes use of the quantum to classical mapping where quantum spin operators $\sigma_i^x$ are mapped to classical spin variables $\sigma_i = \pm1$ with $\sigma^x_i |\sigma_i\rangle = \sigma_i |\sigma_i\rangle$. The path integral for the spins' density matrix then involves a sum over all configurations of the ket and bra states, which we denote by $\bsigma=\{\sigma_i(t)\}$ and $\bsigma'=\{\sigma_i'(t)\}$, respectively. 
Performing the Gaussian integral over the bosonic modes, 
one obtains  
the Feynman-Vernon influence functional  
\begin{align}\label{eq:FV_intro}
    \begin{split}
    {\mathscr F} [\bsigma,\bsigma'] 
    = \exp\Big[-\frac{1}{N}\sum_{ijtt'} & \,\frac{1}{2}\,C^r_{ij} (t,t') \,\tilde\eta_i(t)\tilde \eta_j(t') \\
    +&  \,\,i \chi_{ij}^r (t,t')\,\tilde\eta_i(t)\eta_j(t')    \Big]
    \end{split}
\end{align}
where
\(
    \eta = (\sigma+\sigma')/2
\)
and 
\(
\tilde \eta = (\sigma- \sigma')/2
\)
and the time and spin indices are implicit. The kernels $C^r$ and $\chi^r$ denote the real part of the correlators $C$ and $\chi$ defined as 
\begin{align}\label{eq:chi_and_C}
    \begin{split}
         C_{ij}(t,t') &= \sum_{\alpha} g_{\alpha i}g_{\alpha j} \left\langle [a_\alpha(t) , a^\dagger_\alpha(t')]_+ \right\rangle_B \\
        i\chi_{ij} (t,t') &= \Theta(t-t')\sum_{\alpha} g_{\alpha i}g_{\alpha j}\left\langle [a_\alpha(t) , a^\dagger_\alpha(t')]_- \right\rangle_B
    \end{split}
\end{align}
These expressions are given for \textit{free} bosons in the absence of the coupling to spins (indicated by subscript $B$), and can be explicitly computed.
They have very different interpretations though: while $C_{ij}$ involves the anticommutator and characterizes symmetrized correlations, the function $\chi_{ij}$ encodes the causal response of the bath. Moreover, $C^{r}_{ij} (t,t')$, viewed as a matrix with the rows $it$ and columns $jt'$, is both positive and symmetric.
Using this fact, 
one can make a Hubbard-Stratonovich transformation as 
\begin{equation}\label{eq:HS_model}
    \overline{\exp\Big[\sum_{it}i k_i(t) \tilde\eta_i(t) \Big]} = \exp\Big[-\frac{1}{2N}\sum_{ijtt'} C^r_{ij} (t,t') \tilde\eta_i(t)\tilde \eta_j(t')   \Big]
\end{equation}
where we have introduced a Gaussian distributed (real-valued) field $k_i$ whose correlations are given by $\overline{k_i(t) k_j(t')}=\frac{1}{N} C^r_{ij}(t,t')$; the overline indicates an average with respect to the Gaussian distribution. 
$k_i$ can be then viewed as a longitudinal, though stochastic, field acting on spin $i$. Remarkably, the spin-spin coupling via $C_{ij}$ can be traded for a stochastic sampling while crucially maintaining a unitary evolution for each realization of the stochastic field. 

The causal response function however poses a significant challenge, and is indeed the root of 
the dynamical sign problem in Monte-Carlo type simulations \cite{MAKAROV1994482,Egger1994}. Causality implies that $\chi_{ij}(t,t')$ is not symmetric  
or positive (as a matrix). 
A special case arises for a single spin coupled to an Ohmic bath where $\chi(t,t')$, approximated as a delta function, can be integrated into the dynamics generator \cite{Peter_Stochastic_Schrodinger_Equation,LeHur_2010,lesovik2002dynamics,LeHur_2014,kamar_spin_boson_model_stochastic_SCH}; 
however, this approach is further limited to weak coupling and a large bath; see also related work \cite{Stockburger1998,stockburger1999stochastic}. 
More generally, one can take a similar approach to \cref{eq:HS_model} by applying another Hubbard-Stratonovich transformation to the term involving $\chi$. With suitable correlations between the two stochastic fields, the spins are decoupled at the expense of stochastic sampling. 
But, this comes at the cost of complex-valued fields and non-unitary dynamics. The stochastic evolution then leads to unphysical states that are, among other things, not hermitian. 
Different variations of this approach have been applied, with some success, to models ranging from the prototypical spin-boson model 
(single spin coupled to an infinite bath) 
\cite{Stockburger_2002,STOCKBURGER_2004,Koch_2008,LeHur_2016}, to quantum spin chains
\cite{Hogan_2004,Galitski_2011,Gritsev_2013,De_Nicola_2020,Begg_2020}. Note that the latter emerge in a limit of \cref{eq:hamiltonian} just as the spin-motion coupling gives rise to Ising interactions in trapped ions \cite{Monroe_2021}. Such stochastic approaches however suffer from unstable solutions or slow convergence although different sampling improves their behavior \cite{STOCKBURGER_2004,Koch_2008}.
At a fundamental level, it is not entirely satisfactory that the response function $\chi_{ij}(t,t')$ is treated in a similar fashion as $C_{ij}(t,t')$ via a Hubbard-Stratonovich transformation, hiding the fact that $\chi_{ij}$ is inherently a causal object. 
Furthermore, this approach as well as alternative stochastic methods typically assume that the initial state is factorized and bosons (or, the \textit{bath}) are initially thermal \cite{deVega_2017}.

\section{Main results}\label{sec:main_results}
In this paper, we take a fundamentally different approach to decoupling the spins (both in time $t$ and the spin index $i$) that is manifestly causal. 
A first hint is that the response function contains information about dissipation. Indeed, Markovian dissipation plays an important role in our approach. 
Departing from the Feynman-Vernon influence functional, we develop a hybrid quantum-classical approach to the spin-boson model where bosonic operators are treated using functional integral methods while spins are evolved quantum mechanically. Furthermore, our approach allows us to consider a general initial state that is not necessarily factorized or thermal. Lastly, we present a uniform approach to models that are typically treated with different techniques.

\begin{figure*}[t]
\centering
\includegraphics[scale=0.88]{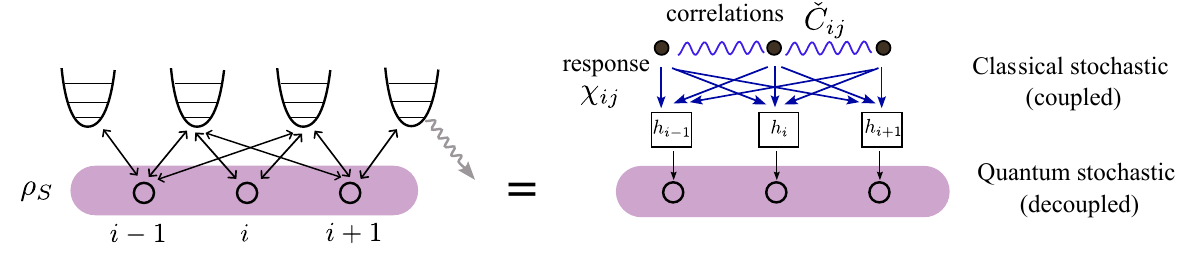}
\caption{(\textbf{Left}) The schematic representation of the spin-boson model where each spin (an open circle) interacts with one or many bosonic modes (characterized by infinite towers of states). The bosonic modes are subject to Markovian loss (wavy arrow). The main object of interest is $\rho_S(t)$ representing the time-dependent reduced density matrix of the spins. (\textbf{Right})~In a hybrid quantum-classical stochastic approach, the solution to a classical stochastic equation mimicking the bosonic modes feeds into the quantum stochastic evolution of the spins. 
Specifically, the coupling to bosons 
is captured by a local longitudinal field $h_i(t)$ that is linearly dependent on both white and colored noise (collectively represented by solid circles), and under which each spin evolves independently.  
The field $h_i(t)$ is however complex valued: the real part of $h_i$ can be viewed as a physical, though time-dependent, field comprising a term that is causally dependent on white noise 
via the response function $\chi_{ij}(t,t')$ plus colored noise whose correlations are given by the \textit{modified} correlation function $\check C_{ij}(t,t')$;
on the other hand, the imaginary part of $h_i$ is directly proportional to white noise, and leads to an unphysical evolution for a given noise realization. Still, our causal approach ensures the density matrix remains hermitian for each realization.
The average over many realizations gives the physical state of the spins and produces its nontrivial correlations. 
See the text for details. }
\label{fig:schematic}
\end{figure*}

We first start with a model where $N$ spins are coupled to a single bosonic mode ($M=1$); we denote the bosonic parameters by $\omega,\kappa$, and the coupling to each spin by $g_i$. 
We show that each spin can be evolved independently while the coupling to the bosonic mode can be mimicked by a classical stochastic field that linearly depends on white noise. 
To this end, we first introduce for each spin a complex-valued white noise $\xi_i(t)$ satisfying
\begin{equation}
    \overline{\xi_i(t)\bar \xi_j(t')} = \kappa\delta_{ij} \delta(t-t')
\end{equation}
We further define the associated field $\underline{\psi}$ via
\begin{equation}
    (i\partial_t - \omega+i\kappa)\underline\psi = \frac{1}{\sqrt{N}}\sum_i\xi_i(t)
\end{equation}
together with the initial value $\underline \psi(t=0)$ that is sampled from the  Wigner distribution function describing the initial (not necessarily thermal) state of the bosonic mode. For simplicity, we assume that the spins and bosons are initially uncorrelated, but it is straightforward to relax this assumption.
We now state our main result (for $M=1$):
for a given realization, the evolution of each spin is decoupled from the rest, and is governed by 
\begin{equation}\label{eq:main_1}
    \frac{d}{dt}\varrho_i = {\cal L}_i(\varrho_i) -i \left[h_i(t) \sigma^x_i  \varrho_i - \bar h_i(t)\varrho_i \sigma^x_i\right]
\end{equation}
where $\varrho_i$ describes the (un-averaged) density matrix for spin $i$, and the longitudinal field $h_i$ is defined from  white noise $\xi_i$ and the associated classical stochastic field $\underline\psi$ as 
\begin{equation}
    h_i = \frac{g_i}{2} (\underline\psi+\underline{\bar\psi}) +i\frac{g_i}{4\kappa} (\xi_i+\bar \xi_i) 
\end{equation}
Finally, summing over many realizations, one obtains the physical density matrix of the spins as 
\begin{equation}
    \rho_S(t) = \overline{\otimes_i \varrho_i(t)}
\end{equation}
where the overline indicates an average over both noise and initial conditions. 
This constitutes a hybrid quantum-classical approach where the solution of a classical stochastic equation feeds into the quantum stochastic evolution of the spin. Remarkably, both classical and quantum stochastic equations are Markovian. This should be contrasted with a Hubbard-Stratonovich transformation
similar to \cref{eq:HS_model} that would introduce colored noise. 
More importantly, causality is manifest through the dependence of the classical stochastic field on noise [$\delta \underline\psi(t)/\delta\xi_i(t')$ gives the response function]. 
As a result, the density matrix remains hermitian even for a single noise realization. Again, these features are to be contrasted with a naive Hubbard-Stratonovich transformation of the term involving the response function. 
Still, the imaginary part of $h_i$ leads to an unphysical evolution, and thus $\varrho$ is not guaranteed to be trace-1 or positive for single trajectories. 
We also remark that \cref{eq:main_1} comes with the multiplicative noise and is given in the It\^{o} sense. 
Notably, the stochastic solutions to \cref{eq:main_1} always exist unlike stochastic methods based on the positive $P$ representation 
\cite{gardiner2004quantum}.

A generalization to $M$ bosonic modes is rather straightforward. A first approach is to generalize our main results above by introducing $M$ classical fields that satisfy linear stochastic equations with $M\times N$ noise variables. Each spin should be then evolved under a stochastic longitudinal field that depends linearly on these fields and noise variables. This approach becomes more expensive for large (or infinite) $M$. Instead, we can use a trick that reduces $M\times N$ white noise variables to just $N$ (again denoted by $\xi_i$), but we must introduce colored noise (dubbed $x_i$) that captures the remaining redundancy. 
In essence, this is similar to the Hubbard-Stratonovich transformation in \cref{eq:HS_model}, while preserving causality explicitly. We further assume that the initial state is factorized and bosonic modes are initially in their vacuum state, but we later show that these assumptions can be easily relaxed. 
More precisely, we evolve each spin under a longitudinal field $h_i$ given by 
\begin{align}
    \begin{split}
         h_i (t)  = &  \frac{1}{2N} \sum_{j=1}^N \frac{1}{\gamma_j}\int_0^\infty dt' \chi_{ij}(t,t') \xi_j(t')+ c.c. \\
         & +\frac{1}{2}(x_i (t)  +\bar x_i(t))  +  \frac{i}{2}\big( \xi_i(t)  +\bar\xi_i(t)\big)
    \end{split}
\end{align}
with the complex-valued noise variables $\xi_i(t)$ and $x_i(t)$: $\xi_i$ is white noise with the correlations 
\begin{equation}
    \overline{ \xi_i(t) \bar\xi_j(t') } = \gamma_i\delta_{ij}\delta(t-t'), \quad \gamma_i =\sum_\alpha \frac{g_{\alpha i}^2}{4\kappa_\alpha}
\end{equation}
and $x_i$ is Gaussian-distributed colored noise whose correlations are given by 
\begin{equation}
    \overline{ x_i(t) \bar x_j(t')} = \frac{1}{N} \check C_{ij}(t,t') 
\end{equation}
where the \textit{modified} correlations $\check C_{ij}(t,t')$ is given by
\begin{align}
        \begin{split}
         \check C_{ij}(t,t') &=  C_{ij}(t,t') \\
        &\,\,\,-\frac{1}{N}\sum_{l=1}^N \frac{1}{\gamma_l}\int_0^\infty\!\!\! dt''  \chi_{il} (t,t'') \bar\chi_{jl} (t',t'')
        \end{split}
\end{align}
The kernels $C$ and $\chi$ are defined according to \cref{eq:chi_and_C}.
One can show that $\check C_{ij}(t,t')$ is positive as a matrix (in the basis $it,jt'$).
Again, the longitudinal field finds an  imaginary component proportional to white noise, hence a non-unitary evolution; however, the density matrix remains hermitian for each trajectory. 
Finally, summing over many realizations, one obtains the physical density matrix.
While each noise realization gives rise to  decoupled, and possibly unphysical, states, the noise average should yield the physical state of the spins and produce their nontrivial correlations. For a schematic representation of the model and our stochastic approach, see \cref{fig:schematic}.

\section{Spin coupled to 
cavity mode $N=M=1$}\label{sec:N=M=1}

In this section, we consider the Rabi model as the simplest nontrivial case of \cref{eq:hamiltonian,eq:liouvillian} with $N=M=1$;
to simplify notation, we take $\omega_\alpha \to \omega$, $\kappa_\alpha\to \kappa$, and $g_{\alpha i}\to g$: 
\begin{equation}\label{eq:H_11}
    H_{1, 1}= \frac{\Delta}{2} \sigma^z+ \omega a^\dagger a  +   \frac{g}{2} (a+a^\dagger)   \sigma^x
\end{equation}
We assume that the system is initially in a factorized state, $\rho(t=0) = \rho_{S}(0)\otimes \rho_{B}(0)$; we shall relax this assumption later. 
Conveniently, we can first ``vectorize'' the density matrix $\rho \to |\rho\rrangle$
such that the Liouvillian  $\cal L$ becomes
a (non-Hermitian) matrix $\mathbb L$ acting on the vectorized state. More explicitly, we map $O|i\rangle \langle j| O' \to O|i\rangle \otimes  {O'}^T |j\rangle$ where the element $|i\rangle \langle j|$ of the density matrix is mapped to the vector 
$|i\rangle \otimes |j\rangle =|i\rangle|j\rangle \equiv |ij\rrangle$.
In the absence of the spin-boson coupling ($g=0$) and starting from the initial state $|\rho_S(0)\rrangle$, the spin simply evolves under a Schr\"{o}dinger-like equation as 
\begin{equation}\nonumber
    \frac{d}{dt}|\rho_S\rrangle = \mathbb L_S |\rho_S\rrangle 
\end{equation}
where the matrix  $\mathbb L_{S}$ 
denotes the spin-only dynamics generator corresponding to ${\cal L}_S$. 
For example, the spin-only Hamiltonian $H_S$ corresponds to ${\mathbb L_S = -i (H_S\otimes I_S- I_S\otimes H_S)}$ with $I_S$ the identity matrix for the spin; spin spontaneous emission can be included in $\mathbb L_S$ as well. For a time-independent $\mathbb L_S$, the state at time $t$ is given by 
\( 
    |\rho_S(t)\rrangle =\exp(\mathbb L_S t)|\rho_S(0)\rrangle.
\) 

More generally, the spin-boson interaction entangles 
the spin and the bosonic mode, and the evolved state can no longer be written in a factorized fashion. 
To tackle this problem, we adopt a hybrid quantum-classical approach, as explained below. The bosonic part of the dynamics, absent the coupling, is simply that of a (damped) quantum harmonic oscillator. The coupling too is linear in the bosonic variable, although a strong coupling renders the dynamics highly nonlinear. However, from a formal perspective, it is straightforward to capture the bosonic part via a functional integral formalism where bosonic operators are mapped to classical phase-space variables and a sum is performed over different classical configurations weighted by a classical action (`classical' refers to a representation of quantum operators as $c$ numbers). In principle, this approach can be extended to the spin, using various representations of spin in terms of coherent states \cite{SachdevBook} or Majorana fermions \cite{BEREZIN1977336}; however, we shall instead treat the spin quantum mechanically. 
We keep track of the spin dynamics via the reduced density matrix $\rho_S(t)= \tr_B(\rho(t))$; we stress that the density matrix is not factorizable. 
Utilizing  a combination of path-integral techniques for the bosonic mode \cite{foss-feig_emergent_2017} together with the quantum-to-classical mapping for the spin \cite{Paz_2021_EPL,Paz_2021_exact}, 
we find 
\begin{align}\label{eq:q2c_mapping}
     \begin{split}
     &| \rho_{S}(t)\rrangle =\int {\mathscr D}[\psi, \phi] e^{i {\mathscr S}_B} {\mathscr W}_0(\psi_0) \times   \\
     &\times  {\rm T}_t\, e^{\int_0^t dt' \,  (\mathbb  L_{S}+ \mathbb L_{\rm int}(t'))}|\rho_{S}(0)\rrangle
     \end{split}
 \end{align}
We shall leave the details of the derivation to \cref{sec:derivation_rho}, and just explain the different terms in this expression:
the fields $\psi, \phi$ are the phase-space variables used to map the bosonic operators to \textit{c} numbers, 
and the corresponding action ${\mathscr S}_{B}$ is given by 
\begin{equation}\label{eq:S_B}
    {\mathscr S}_B = \int_0^t dt' \left[ 2\bar \phi\big(i\dot\psi  - \omega \psi+ i\kappa \psi\big) +c.c.  + 4i\kappa \bar\phi\phi\right]
\end{equation}
The function ${\mathscr W}_0(\psi_0)$ denotes the Wigner function corresponding to the initial state of the cavity mode, $\rho_B(0)$. The first line of \cref{eq:q2c_mapping} then involves a (functional) integration over both $\psi, \phi$ weighted by the exponential of the action ${\mathscr S}_B$ as well as the Wigner function ${\mathscr W}_0(\psi_0)$ corresponding to the initial state.
Finally, the second line of \cref{eq:q2c_mapping} is the  time-ordered product (enforced by the time ordering operator ${\rm T}_{t}$) of the evolution operator that involves, besides the spin-only part, $\mathbb L_S$, an interaction-induced matrix $\mathbb L_{\rm int}(t)$ due to the coupling to the bosonic mode. The latter matrix takes the form
\begin{equation}\label{eq:tilde_L_2}
    \mathbb L_{\rm int}(t) = i g \left(\psi + \bar \psi\right) \mathbb S +ig\left(\phi + \bar \phi\right) \mathbb T
\end{equation}
where the time dependence of the fields $\psi(t), \phi(t)$ are  implicit, and the matrices $\mathbb S, \mathbb T$ are defined as 
\begin{equation}\label{eq:T_c/q}
\begin{split}
    {\mathbb S} &= -\frac{1}{2} \left(\sigma^x\otimes I_S - I_S \otimes \sigma^x\right) \\
    {\mathbb T} &= -\frac{1}{2} \left(\sigma^x\otimes I_S + I_S \otimes \sigma^x\right)
\end{split}
\end{equation}
One can see that the term proportional to $\mathbb S$ in \cref{eq:tilde_L_2} can be interpreted as a \textit{classical}, though time-dependent, longitudinal magnetic field, while the term proportional to $\mathbb T$ does not admit such interpretation. 

It is particularly convenient to work in a basis that diagonalizes the spin-boson interaction. To this end, we first define $|\sigma\rangle=|\pm\rangle$ as eigenstates of the operator $\sigma^x$, that is,   
$\sigma^x|\sigma\rangle= \sigma|\sigma\rangle$. 
A matrix (such as $\mathbb L_S$) can be then represented in a basis spanned by $|\sigma\sigma'\rrangle \in \left\{|++\rrangle, |+-\rrangle,|-+\rrangle,|--\rrangle\right\}$. 
In this basis, the matrices $\mathbb S, \mathbb T$ become diagonal and take a simple form, 
\begin{align}\label{eq:Tc/q_diagonal}
    \mathbb{S}  = \,{\rm diag}\{0,-1,1,0\}\,, \quad \mathbb{T}  = \,{\rm diag}\{-1,0,0,1\}
\end{align}
and the interaction matrix becomes 
\begin{align}
    \mathbb L_{\rm int}(t)=2ig
    \renewcommand*{\arraystretch}{1}
    \begin{pmatrix}
    -(\phi +\bar\phi) & 0  & 0 & 0 \\  
    0 & -(\psi +\bar\psi)   & 0 & 0 \\ 
    0 & 0 & \psi +\bar\psi  &  0 \\ 
    0 & 0 & 0 & \phi +\bar\phi
    \end{pmatrix} \label{eq:tilde_L}
\end{align}

In short, \cref{eq:q2c_mapping} defines a hybrid approach where the spin is still explicitly quantum mechanical while the bosonic operators are traded in for the phase-space variables $\psi(t),\phi(t)$. The resulting functional integral is however rather formal and is of little practical use because of the sign problem, that is, it involves complex-valued weights and is not amenable to sampling via a Monte-Carlo type of approach. 

Here, we take a different approach utilizing Markovian dissipation ($\kappa\ne 0$). As a first step, we use a standard trick that converts stochastic Langevin equations to a path integral and vice versa \cite{KamenevBook}. To this end, we write the last term in the action in \cref{eq:S_B}, sometimes referred to as the ``quantum noise'', in terms of a noise field $\xi(t)$ using a Hubbard-Stratonovich transformation: 
\begin{equation}\label{eq:HS}
    e^{-4\kappa \int_t \bar \phi\phi} = \int {\mathscr D}[\xi] e^{-\int_t \bar \xi \xi  /\kappa- 2i\int_t  (\bar \xi \phi+ \xi \bar \phi)}
\end{equation}
As a warm up, let us first consider $g=0$, so that there is no spin-boson interaction (specifically, $\mathbb L_{\rm int}=0$). 
This is a trivial exercise (the first line of \cref{eq:q2c_mapping} just yields 1), but it sets the stage for later. 
With the above transformation and in the absence of the spin-boson coupling, the field $\phi$ only appears linearly in the action, and its (path) integral yields a delta function which enforces a stochastic equation of motion,
\begin{equation}\label{eq:langevin}
    i\dot{\underline{\psi}}  - \omega \underline{\psi}+ i\kappa \underline{\psi} = \xi (t)
\end{equation}
where $\xi(t)$ can be viewed as white noise with the correlations
\begin{equation}\label{eq:white_noise}
    \overline{\xi(t) \bar\xi(t')}  = \kappa \delta(t-t')
\end{equation}
and $\overline{\xi(t) \xi(t')}=0$. The above equations should be supplemented with the initial condition $\underline{\psi}(t=0)=\psi_0$ which is drawn from the Wigner distribution function ${\mathscr W}_0(\psi_0)$. The underline emphasizes that $\underline\psi$ is not a free field, and is completely fixed  by $\xi(t)$ and $\psi_0$.

Turning on the spin-boson interaction ($g\ne 0$), 
the field $\phi$ also appears in the time-ordered product in the second line of \cref{eq:q2c_mapping} through $\mathbb L_{\rm int}$ which is explicitly defined in \cref{eq:tilde_L_2}. 
Therefore, one cannot immediately integrate over $\phi$ in the same fashion as described above.
The trick is to instead write $\phi$ in the time-ordered product
as a (functional) derivative with respect to $\xi$, 
\begin{equation}
    {\rm T}_t  e^{i g \int_t (\phi+\bar \phi) \mathbb T + \cdots}\longrightarrow {\rm T}_t  e^{-\frac{g}{2} \int_t \mathbb T ({\delta}/\delta {\bar\xi}+\delta/\delta{\xi})+\cdots}  \nonumber
\end{equation}
acting on $\exp(-2i\int_t \bar \xi\phi + \xi \bar\phi)$ introduced in the Hubbard-Stratonovich transformation in \cref{eq:HS}; the dots represent the remaining terms in $\mathbb L + \mathbb L_{\rm int}$ which are dropped for ease of notation. Notice that the above operation  simply induces a shift in the noise variable $\xi \to \xi - \frac{g}{2} \mathbb T$ where the first term is now understood to be proportional to the identity matrix $\mathbb I={\rm diag}\{1,1,1,1\}$ in the vectorized space. 
An integration by parts  allows us to put the partial derivatives on the noise Gaussian distribution (functional), $\exp (-\int_t \bar\xi \xi /\kappa)$, which can be then explicitly evaluated by applying the inverse shift.
The net effect of this procedure is to replace 
the last term in \cref{eq:tilde_L_2} as
\begin{align}\label{eq:phi_to_xi}
    i g(\phi+\bar\phi) \mathbb T  \longrightarrow -\frac{g}{2\kappa}(\xi+\bar\xi) \mathbb T - \frac{g^2}{4\kappa}\mathbb T^2
\end{align}
With this transformation, the field $\phi$ now appears  only in the action ${\mathscr S}_B$, and the path integral over this field can be explicitly done, which in turn 
constrains $\psi$ via \cref{eq:langevin}.
These steps can be rigorously justified by discretizing time in the functional integral and Trotterizing the evolution operator that appears in the time-ordered product. We leave the details to \cref{sec:derivation_rho_xi}, and just report the final result: 
\begin{equation}\label{eq:q2c_4}
    \begin{split}
    | \rho_{S}(t)\rrangle = &\int {\mathscr D}[\xi] e^{-\int_t \bar \xi \xi/\kappa} \int d^2\psi_0{\mathscr W}_0(\psi_0) \times \\
    &\times  {\rm T}_t\, e^{\int_0^t dt' \,  {\mathbb  K}(t')}|\rho_{S}(0)\rrangle
    \end{split}
\end{equation}
where the matrix $\mathbb K(t)$ is given by 
\begin{equation}\label{eq:K-tilde}
    {\mathbb K} (t) = \mathbb L_S + ig (\underline\psi + \underline{\bar\psi})\mathbb S - \frac{g}{2\kappa} (\xi+\bar \xi) \mathbb T-\frac{g^2}{4\kappa} \mathbb T^2
\end{equation}
and depends on time implicitly through the noise $\xi$ and the associated field $\underline{\psi}$.
Notice that the functional integral over $\psi, \phi$ is now replaced by an integral over noise and the initial Wigner distribution function in \cref{eq:q2c_4}. Most importantly, the weight of the  functional integral is now positive (at least when the Wigner function is positive). 

Next, we define $|\rho_S(t)\rrangle_{\xi}$ as  the time-ordered product in the second line of \cref{eq:q2c_4}:
\begin{equation}\label{eq:rho_xi}
    |\rho_S(t)\rrangle_{\xi} \equiv  {\rm T}_t\, e^{\int_0^t dt' \,  {\mathbb  K}(t')}|\rho_{S}(0)\rrangle
\end{equation}
for a given noise realization $\xi(t)$ and the initial value $\psi_0$; the dependence of $|\rho_S(t)\rrangle_\xi$ on $\psi_0$ is made implicit for ease of notation. The full density matrix is obtained by averaging  over different realizations, which is again free of the sign problem. 
It follows from \cref{eq:rho_xi} that the dynamics can be written in terms of a generator, i.e., via an equation that is local in time:
\begin{equation}\label{eq:stoch-L-final-1}
    \frac{d}{dt} |\rho_S\rrangle_{\xi} = \mathbb K^{\rm I}(t) |\rho_S\rrangle_{\xi}
\end{equation}
where $\mathbb K^{\rm I} (t) ={\mathbb K}(t) +\frac{g^2}{4\kappa}\mathbb T^2$. \Cref{eq:stoch-L-final-1} is a stochastic equation with multiplicative white noise, and is given in the sense of It\^{o}. In fact, the difference between $\mathbb K$ and $\mathbb K^{\rm I}$ follows from the It\^{o} rule; a careful derivation is provided in \cref{sec:derivation_rho_xi}.
Notice that the extra term in the definition of  $\mathbb K^{\rm I}$ 
just cancels out against the last term in \cref{eq:K-tilde}, hence the resulting simple equation $\mathbb K^{\rm I}=\mathbb L_S + ig (\underline\psi + \underline{\bar\psi})\mathbb S - \frac{g}{2\kappa} (\xi+\bar \xi) \mathbb T$. Interestingly, this equation can be identified simply by substituting $\phi \to -i\xi/\kappa$ rather than \cref{eq:phi_to_xi}. 
Adopting the basis defined by  \cref{eq:Tc/q_diagonal}, the dynamics generator is explicitly given by 
\begin{widetext}
    \begin{align}\label{eq:K_S_Ito}
    \mathbb K^{\rm I}(t)  
    = \mathbb L_S + 
    \renewcommand*{\arraystretch}{1}
    \begin{pmatrix}
     \frac{g}{2\kappa} (\xi+\bar\xi) & 0 & 0 & 0 \\ 
    0 &  - i g (\underline{\psi}+\underline{\bar\psi})  & 0 & 0\\ 
    0 & 0 &  ig (\underline{\psi}+\underline{\bar\psi}) & 0 \\ 
    0 & 0 & 0 &  -\frac{g}{2\kappa}(\xi+\bar\xi)
    \end{pmatrix}
    \end{align}
\end{widetext}
Calculating the state $|\rho_S(t)\rrangle_{\xi}$ for a given realization from \cref{eq:stoch-L-final-1,eq:K_S_Ito}, we
can then find the full time-dependent state by averaging over all realizations,
\begin{equation}\label{eq:stoch-L-final-2}
    |\rho_S(t)\rrangle =
    \overline{|\rho_S(t)\rrangle_{\xi}}
\end{equation}
where, in a slight abuse of notation, the overline denotes the average over both the noise as well as the initial conditions,
\begin{equation}\label{eq:avg_xi_ps0}
    \overline{\bigskip\,\cdots\,} = \int {\mathscr D}[\xi] e^{-\int_t \bar \xi \xi/\kappa} \int d^2\psi_0{\mathscr W}_0(\psi_0) \cdots
\end{equation}
We remark in passing that the integration over the Wigner function representing the initial state bears resemblance to the truncated Wigner approximation \cite{POLKOVNIKOV20101790}; however, the stochastic approach presented here is \textit{exact}.

The above equations are among the main results of this paper. 
These equations feature  several important, and immediately useful, properties. 
Here, we shall summarize and further highlight these points, and furthermore enlist other important features of these equations. 

\begin{itemize}[wide, labelwidth=!, labelindent=0pt, leftmargin=0pt]
    \item \textbf{Non-factorized initial state.} Our treatment can be readily generalized to a non-factorized initial state $\rho(0)$ via the substitution 
    \[{{\mathscr W}_0(\psi_0) \rho_S(0) \to \widetilde{\rho_S}(\psi_0) \equiv\tr_B[\delta_{\rm W}(\psi_0-a)\rho(0)]}
    \] where the Weyl-ordered delta function is defined as $\delta_{\rm W}(\psi-a)=\int \frac{d^2\phi}{\pi^2}\exp[\bar \phi(\psi-a)-\phi(\bar\psi-a^\dagger)]$.  For a given $\psi_0$, we should then evolve the spin starting from the initial state given by $\widetilde\rho_S(\psi_0)$. \Cref{eq:avg_xi_ps0} is then replaced by an average over $\psi_0$ that is effectively sampled by $\widetilde{\rho_S}(\psi_0)$. For a factorized initial state, we recover the Wigner function ${\mathscr W}_0(\psi_0)\equiv\tr_B[\delta_{\rm W}(\psi_0-a)\rho_B(0)]$ \cite{POLKOVNIKOV20101790,foss-feig_emergent_2017}. 
    
    \item \textbf{Feynman-Vernon influence functional.} 
    For an initial state where the boson is in its vacuum state, the Wigner function is a Gaussian function, ${\mathscr W}_0(\psi_0)=\frac{2}{\pi}\exp(-2|\psi_0|^2)$. In this case, one can show that \cref{eq:q2c_4} directly leads to the influence functional in \cref{eq:FV_intro}; see \cref{sec:FV_1}.
    The perspective afforded by the Feynman-Vernon influence functional is particularly useful in our treatment of $N$ spins coupled to boson(s); see \cref{sec:N>1_M=1}. 
    
    \item \textbf{Sign-problem free at expense of unphysical trajectories.} The original path-integral formulation in \cref{eq:q2c_mapping}, or the Feynman-Vernon formalism, are exact, but they suffer from the dynamical sign problem. On the other hand, the stochastic formulation in \cref{eq:q2c_4} is sign problem free. (A negative value, if any, of the Wigner function corresponding to the initial state is only a  mild exception.)  However, this comes at the expense of unphysical states for each trajectory where the density matrix is not trace-1 or positive. Still, the causal structure of our approach keeps the density matrix hermitian even for a single trajectory; see \cref{sec:sto_master_eqn} for details.

    \item \textbf{It\^{o} vs Stratonovich.} The stochastic differential equation in \cref{eq:stoch-L-final-1} follows the It\^{o} convention \cite{gardiner1985handbook}
    and involves multiplicative noise,
    $d\rho_m = A_m(\rho)dt + b_m(\rho) dW$ where $dW$ is the Wiener increment, and $A_m, b_m$ are linear functions of the (vectorized) density matrix components $\rho_m$ with $m=1,2,3,4$. Specifically, $b_m(\rho)= -(g/\sqrt{2\kappa}) T_m \rho_m$ with $T_m$ the diagonal element of $\mathbb T$ in the basis defined in \cref{eq:Tc/q_diagonal}.  
    Interestingly, one can see that 
    the dynamics in the Stratonovich sense \cite{gardiner1985handbook} is governed by the matrix $\mathbb K$ that appears in the functional integral.

    \item \textbf{Markovian in form;~non-Markovian in essence.} In our approach, we only keep track of the spin dynamics, while we trade in the cavity mode for a classical stochastic field. 
    This approach is nonperturbative, but remarkably the dynamics remains purely local in time.
    Put differently, an exact elimination  of the bosonic operator is possible (well beyond the domain of adiabatic elimination) while maintaining locality in time.

    \item \textbf{Existence of stochastic solutions.} Stochastic differential equations could often lead to unstable solutions (for example, in methods based on the positive $P$-representation \cite{gardiner2004quantum}). Our approach does not suffer from this since a certain \textit{growth} condition (see Ch.~6 of Ref.~\cite{gardiner2004quantum}) is satisfied, which thus guarantees that stochastic solutions exist at all times. 
    An immediate question though is how many trajectories are required for convergence. We study this question in several examples in this work; see for example  \cref{sec:numerics_11}. 
    In general, the convergence improves for larger dissipation and/or smaller coupling. 
    A systematic study of convergence with the number of trajectories is left to future work.

    \item \textbf{Dissipation as computational resource.} Here, we have used Markovian dissipation to trade in the coupling to the bosonic mode for a stochastic sampling that respects causality and ensures hermiticy for each trajectory. 
    In principle, Markovian dissipation could be taken infinitesimal to simulate unitary dynamics; however, the convergence worsens with decreasing dissipation. 
    This is a satisfactory feature as one expects dissipation, responsible for the quantum-classical crossover, renders the dynamics more amenable to a numerical simulation. In contrast, MPS-based methods, for example, become more complex when dealing with dissipation \cite{Vidal_Mixed_State_MPS,Verstraete_MPS_Open_Quantum_Systems,Jens_Eisert_Positive_Tensor_Open_Quantum_System,Weimer_Open_Quantum_System_Method_Review}. Our approach thus provides a concrete framework where Markovian dissipation can be used as a computational resource; see also \cite{propp2023decoherence}.
    
\end{itemize}

\begin{figure*}[t]
{
\includegraphics[scale=.45
]{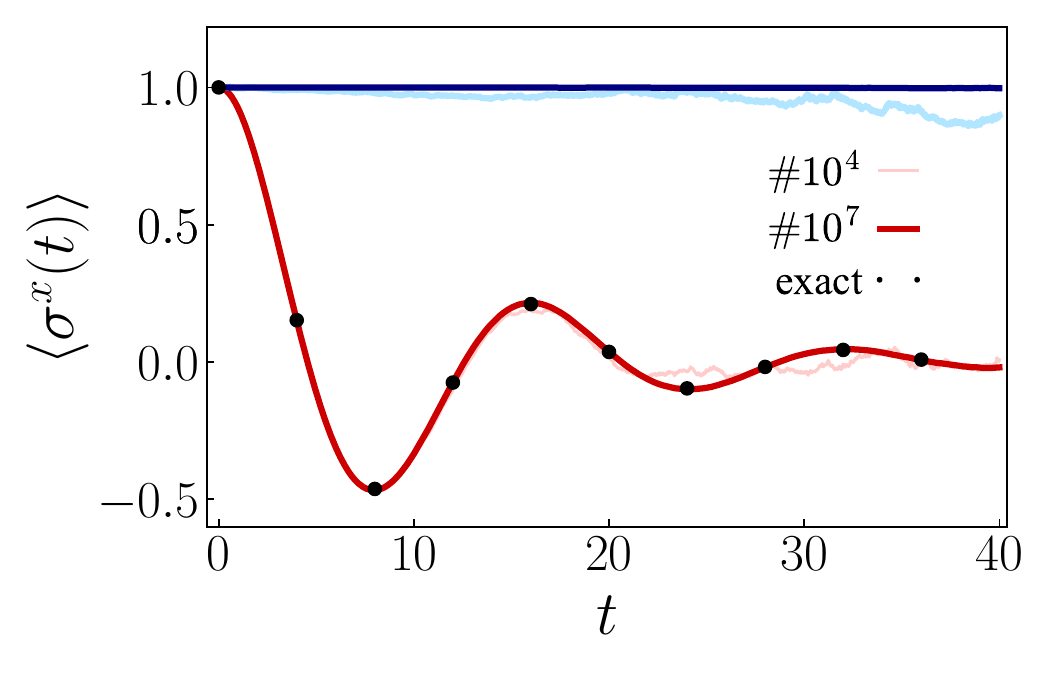}
}
\hspace{.5cm}
{
\includegraphics[scale=.45]{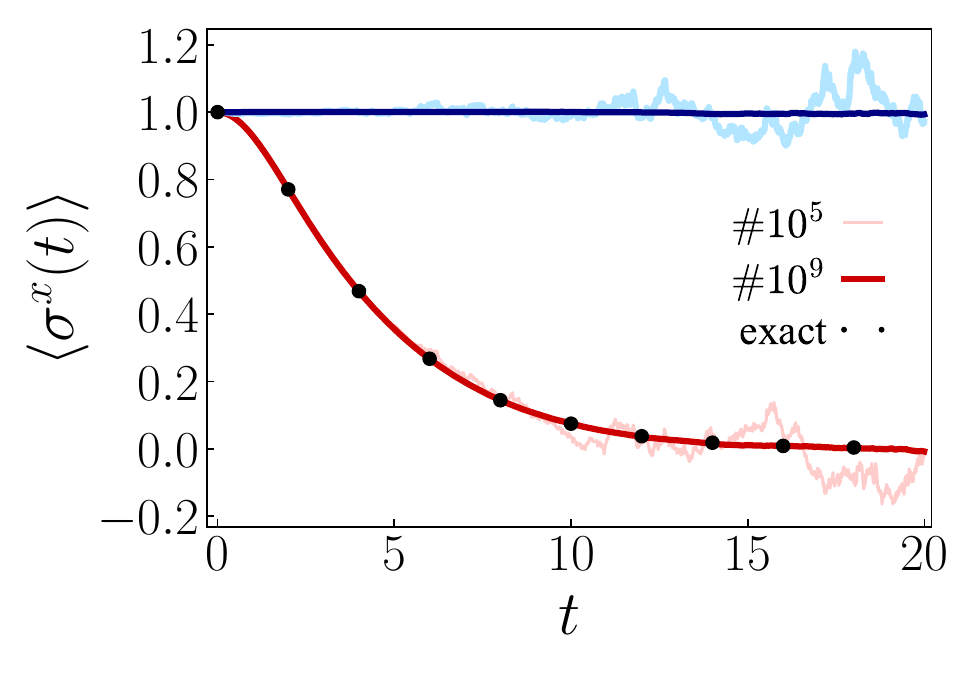}
}
\caption{Stochastic evolution of $\langle \sigma^x(t)\rangle$ starting from an initial state where the spin points along the positive $x$ direction and the boson is in its vacuum state; here, $\omega=\kappa=1, \Delta=0.4$ while $g=0.6, 1.2$ in the left/right panels, respectively. The decay of the overall amplitude is purely due to the coupling to the bosonic mode. (a) Underdamped dynamics at $g=0.6$ for $\#10^4, 10^7$ trajectories. 
The stochastic dynamics with $10^7$ trajectories is in excellent agreement with an exact numerical computation up to $t=40$.  (b) Overdamped dynamics at $g=1.2$ for $\#10^5,10^9$ trajectories. 
Full agreement with an exact numerical computation up to $t=20$ is achieved for $10^9$ different trajectories. In both cases, the averaged trace (see the blue curves) approaches 1 with increasing the number of trajectories. 
\label{fig:N=1}}
\end{figure*}

\subsection{Stochastic master equation}\label{sec:sto_master_eqn}
We can gain further insight by writing the dynamics explicitly for the density matrix, effectively undoing vectorization.
From \cref{eq:q2c_4,eq:K-tilde}, we find that the dynamics in the It\^{o} convention takes the form
\begin{align}\label{eq:master-equation1}
    &\frac{d}{dt}\rho_\xi \equiv  {\cal K}^{\rm I}(\rho_\xi) =  {\cal L}_S(\rho_\xi) -i   \left[h(t)  \sigma^x \rho_\xi -\bar h(t) \rho_\xi\sigma^x\right] \nonumber \\
    &\mbox{with} \quad h(t)=  \frac{g}{2} (\underline\psi+\underline{\bar\psi}) +i\frac{g}{4\kappa} (\xi+\bar \xi) 
\end{align}
for a given noise realization $\xi(t)$ and an initial condition $\underline\psi(t=0)=\psi_0$; for brevity, we have not explicitly shown the dependence of $\rho_\xi$ on $\psi_0$ and dropped the subscript $S$. 
Note that the generator of the full dynamics, ${\cal K}^{\rm I}$, implicitly depends on time through its dependence on $h(t)$. 
The field $h$ can be viewed as a complex-valued longitudinal field:
the real part of $h$ mimics a physical (though stochastic and time-dependent) longitudinal field given by $(g/2) (\underline\psi+\underline{\bar \psi})$.
On the other hand, the imaginary part of $h$ is  proportional to the noise and the corresponding term in the master equation is not of a Hamiltonian form. Still, we can interpret the corresponding dynamics 
as that of a non-hermitian Hamiltonian, $H_h \rho_\xi - \rho_\xi H_h^\dagger$, with $H_h \equiv h(t) \sigma^x$. 
But the evolved state is no longer 
trace 1 or positive. 
The lack of positivity becomes more manifest when writing the master equation in the Stratonovich sense\footnote{Even with ${\cal L}_S=0$, 
\begin{equation}\nonumber
    \rho_\xi(t) \ne U(t)  \rho_0 U^\dagger(t)
\end{equation}
which would otherwise imply positivity; here, ${U(t)={\rm T}_t e^{i\int_0^t dt'  h(t')}}$ (which  is not unitary since $h$ is complex valued). This is because, according to the It\^{o} rule, the evolution due to the non-hermitian Hamiltonian cannot be broken into the evolution of ket and bra states independently. The proper exponentiated evolution matrix is then ${\cal K}$ corresponding to the matrix $\mathbb K$ which is incidentally the dynamics generator in the Stratonovich sense.} by adding the term $-\frac{g^2}{8\kappa}(\sigma^x \rho_\xi \sigma^x + \rho_\xi)$ to the rhs of the first line in \cref{eq:master-equation1}; this term looks like dephasing but with a wrong sign for the \textit{jump} term. It is this sign difference that could lead to negative eigenvalues of the density matrix $\rho_\xi$ for a fixed noise realization.

While the density matrix for each realization is not physical, it remains hermitian even for a single realization since 
\begin{equation}
    ({\cal K}^{\rm I}(\rho_\xi))^\dagger= {\cal K}^{\rm I}(\rho_\xi^\dagger)
\end{equation}
This feature is a direct consequence of our causal treatment, and is particularly convenient for numerical computations as it puts a strong constraint on the form of the  density matrix $\rho_\xi$.
Finally, we note that the average over many realizations 
must yield a physical density matrix that is trace 1 and positive. 
In fact, averaging the master equation in \cref{eq:master-equation1} over noise and using $\overline{\xi(t) \rho_\xi(t)}=0$ from the It\^{o} convention shows that the trace is conserved, $d\overline{\tr(\rho_\xi)}/dt=0$, on average (even before averaging over initial conditions).  
The above properties can be explicitly verified numerically, as we discuss in the next section.

\subsection{Numerical results}\label{sec:numerics_11}

Here, we provide  numerical results for different examples of the model considered in this section. Specifically, we plot $\langle \sigma^x(t)\rangle $ as a function of time starting from  an initial state where the spin is fully polarized along the positive $x$ direction, and the boson is in its vacuum state.
As a representative example, we take $\omega=\kappa= 1, \Delta=0.4$ and consider two characteristic values for the spin-boson coupling, $g=0.6,1.2$; 
all parameters are comparable in order to avoid any fine tuning. To simulate the stochastic dynamics, we have adopted the Euler method \cite{gardiner1985handbook} providing a simple method for solving stochastic equations; more accurate techniques will improve the convergence properties. Finally, we choose the time step $dt=0.01$. 

In \cref{fig:N=1}(a), we plot the dynamics up to $t=40$ for $g=0.6$, and observe underdamped dynamics. Even with a moderate number of trajectories ($10^4$), the dynamics is captured to relatively long times. For a larger number of trajectories ($10^7$), our results perfectly match the exact numerical computation. Additionally, we plot the averaged trace, and observe that this quantity approaches 1 with increasing the number of trajectories.

In \cref{fig:N=1}(b), we plot the dynamics up to $t=20$ for $g=1.2$, and rather 
observe overdamped dynamics. A stronger coupling requires a larger number of trajectories for convergence in a given time interval. While a relatively small number of trajectories ($10^5$) capture the main features of the dynamics, full convergence up to $t=20$ requires a larger number of trajectories (the thick curves are  obtained for $10^9$ trajectories). Again, the trace acts as a proxy for convergence and approaches 1 with increasing the number of trajectories.

\section{$N$ spins coupled to a boson $ M=1$}\label{sec:N>1_M=1}
In this section, we consider  several (or many) spins coupled to a single bosonic mode. The Hamiltonian is now given by 
\begin{equation}
    H_{N, 1}= \omega a^\dagger a +  \frac{\Delta}{2} \sum_{i=1}^N \sigma^z_i +  \frac{1}{\sqrt{N}} (a+a^\dagger)   \sum_{i} \frac{g_{ i}}{2}\sigma^x_i
\end{equation}
Again, the bosonic mode is subject to loss characterized by the Lindblad operator $L=\sqrt{\kappa} a$; spins could be lossy as well. 
A generalization of the main result of the previous section to many spins is straightforward. Let us denote by $|\rho^{(N)}\rrangle$ the (vectorized) reduced density matrix of the $N$ spins upon tracing out the bosonic mode.
The dynamics can be then written as a quantum stochastic evolution with each trajectory evolving as \begin{align}
    \frac{d}{dt}|\rho^{(N)}\rrangle_\xi = \mathbb K^{\rm I}(t)|\rho^{(N)}\rrangle_\xi
\end{align}
where the dynamics generator $\mathbb K^{\rm I}$, now acting on $N$ spins,
is given by
\begin{equation}\label{eq:K_I_N_spin}
    \mathbb K^{\rm I}(t) = \sum_{i=1}^N \left[\mathbb L_i +  i \frac{g_i}{\sqrt{N}}(\underline\psi + \underline{\bar\psi})\mathbb S_i- \frac{g_i}{2\sqrt{N}\kappa} (\xi+\bar \xi) \mathbb T_i\right]
\end{equation}
Here, $\mathbb L_i$ describes the single-body dynamics of the spin $i$, which 
more precisely should be understood as a tensor product with the identity matrix for other spins, $\mathbb I \otimes \cdots \otimes \mathbb L_i \otimes \cdots \mathbb I$. Finally, the white noise $\xi$ and the associated field $\underline\psi$ are defined exactly as before via \cref{eq:langevin,eq:white_noise}. 

In principle, for a given realization of noise and a given initial state, \cref{eq:K_I_N_spin} can be used directly 
to evolve the vectorized density matrix of $N$ spins, i.e., a vector of size $4^N$. The average over many realizations then gives the physical density matrix.
In practice however, even for moderate values of $N$, and certainly in a many-body system, the size of the state 
becomes prohibitively large for any numerical simulation. 
On the other hand, we notice 
that $\mathbb K^{\rm I}$ in \cref{eq:K_I_N_spin} is decoupled among different spins. 
Naively, each spin could be then evolved individually (at least, if  the initial state is factorized) for a given noise realization.
However, this argument is flawed! As a first observation,  
note that $\mathbb K$ appearing in the time ordered product (also, the generator of the dynamics in the Stratonovich sense) 
is given by $\mathbb K = \mathbb K^{\rm I}- \frac{1}{4N\kappa} ( \sum_i g_i\mathbb T_i)^2$ which directly couples the spins. A resolution thus lies 
in the nontrivial form of the It\^{o} chain rule. As a simple example, consider $N=2$ spins and let us assume (to prove the contrary) that the dynamics is decoupled:
\begin{equation}\nonumber
    d\rho_{i m} = A_{im} dt - \frac{g}{2\sqrt{\kappa}} T_m \rho_{i m} dW
\end{equation}
where $i=1,2$ denote the spins and $m=1,2,3, 4$ refer to the component of the corresponding vectorized density matrix. 
The resulting equation for the total density matrix, $\rho^{(2)}_{mn} = \rho_{1m}\rho_{2n}$, is then obtained by applying the It\^{o} chain rule,
\begin{equation}\nonumber 
    d\rho^{(2)}_{mn} = \cdots  + \frac{\partial^2 \big(\rho^{(2)}_{mn}\big)}{ \partial\rho_{1m} \partial\rho_{2n} } d\rho_{1m} d\rho_{2n} =  \cdots + \frac{g^2}{4\kappa} T_m T_n \rho_{mn}^{(2)} dt 
\end{equation}
where the dots refer to single-body terms. 
Notice that the new term on the rhs couples the two spins, while there is no such coupling in \cref{eq:K_I_N_spin}. More generally, for $N$ spins, one finds a coupling between all pairs of spins.
(Interestingly, the Stratonovich dynamics involves a term  of the same form but with an opposite sign.) 
Therefore, regardless of the stochastic rules we adopt, the spins do not seem to be decoupled.\footnote{In essence, this argument is the same as the previous footnote.}
Paradoxically, the absence of any coupling in \cref{eq:K_I_N_spin} for the full (tensor product)
state implies that they are effectively coupled.  
Nonetheless, we can bring the dynamics into a form where individual spins are evolved independently by introducing a different noise variable for each spin. 
We explain this procedure in the following subsection.

\subsection{Decoupling spins}\label{sec:decoupling}
In this subsection, we provide a recipe for decoupling the spins. For convenience, we assume that the initial state of spins is a product state (in the vectorized space), i.e., $|\rho^N(0)\rrangle= \otimes_i |\rho_i(0)\rrangle$; this assumption can be relaxed simply by writing the initial state of spins as a superposition (again, in the vectorized sense) of product states. The proof follows from a variation of the Feynman-Vernon influence functional, which we leave to \cref{sec:FV_2}. The main trick is to introduce uncorrelated fictitious noise variables $\xi_i$ with $i=1,\cdots, N$, i.e., 
\begin{equation}\label{eq:xi_i}
    \langle \xi_i(t) \bar\xi_j(t')\rangle =\kappa\delta(t-t')\delta_{ij}
\end{equation}
and define the field $\underline\psi(t)$ as 
\begin{equation}\label{eq:G_inverse}
    (i\partial_t -\omega + i\kappa)\underline\psi(t) =\frac{1}{\sqrt{N}} \sum_j \xi_j(t)
\end{equation}
For a given noise realization $\bxi = \{\xi_i\}$ and a given initial state, we can write the state in a factorized form at all times,
\begin{equation}
    |\rho^N(t)\rrangle_{\bxi} = \otimes_{i=1}^N  |\rho_i(t)\rrangle_\bxi
\end{equation}
where each spin is evolved as
\begin{equation}
    \frac{d}{dt}|\rho_i\rrangle_\bxi = \mathbb K_i^{\rm I}(t)|\rho_i\rrangle_\bxi
\end{equation}
and the generator of the dynamics is given by 
\begin{equation}
    \mathbb K^{\rm I}_i (t) = \mathbb L_i +  i \frac{g_i}{\sqrt{N}}(\underline\psi + \underline{\bar\psi})\mathbb S_i- \frac{g_i}{2\kappa} (\xi_i+\bar \xi_i) \mathbb T_i 
\end{equation}
Notice that while the same field $\underline\psi(t)$ is coupled to all the spins, each spin $i$ is subject to its own noise $\xi_i(t)$; of course, the dynamics of a given spin still depends on all the noise variables through its dependence on $\underline{\psi}$.
We thus find that the evolution of each spin is given in a similar fashion as that of a single spin in \cref{eq:K_S_Ito} only with the modification $\underline\psi \to  \underline\psi/\sqrt{N}$, $\xi \to \xi_i$, and $g\to g_i$ for spin $i$ together with the noise correlations and the stochastic equation of motion in \cref{eq:G_inverse,eq:xi_i}. Finally, the physical state  is given by the average over noise and the initial state
\begin{equation}\label{eq:rho_N}
    |\rho^{(N)}\rrangle  =  \overline{|\rho^{(N)}\rrangle_\bxi}
\end{equation}
which, again in a slight abuse of notation, the overline indicates the average over all the noise variables $\bxi$ as well as the initial conditions,
\begin{equation}
    \overline{\bigskip\,\cdots\,} = \int {\mathscr D}[\bxi] e^{-\sum_i\int_t \bar \xi_i \xi_i/\kappa} \int d^2\psi_0{\mathscr W}_0(\psi_0) \cdots
\end{equation}
 
As they evolve, the spins  form nontrivial correlations or become entangled.
In our approach, the sum over different realizations effectively mimics the quantum (and statistical) correlations between the spins. 
In principle, this leads to an exponential reduction from a state of size $4^N$ to $N$ vectors of size 4. Of course, the number of trajectories required for convergence could limit the applicability of this method. 
In practice, we do not need to keep track of the full state in \cref{eq:rho_N} if we are interested in expectation values of local operators, two-, or $n-$point correlation functions. For example, the expectation value of an operator $O_i$ acting on spin $i$ is simply given by 
\begin{equation}
    \langle O_i(t) \rangle =\overline{ \llangle O^T_i|\rho_i(t)\rrangle_\bxi  \prod_{l\ne i} \llangle I_l|\rho_l(t)\rrangle_\bxi}
\end{equation}
and the correlation function between two spins is given by 
\begin{equation}
    \langle O_i(t) O_j(t) \rangle =\overline{ \llangle O^T_i|\rho_i(t)\rrangle_\bxi\llangle O^T_j|\rho_j(t)\rrangle_\bxi  \prod_{l\ne i,j} \llangle I_l|\rho_l(t)\rrangle_\bxi}
\end{equation}
and similarly for higher $n$-point correlation functions. Similarly, we can find a simple expression 
for the reduced density matrix for a subset of spins. For example, the reduced density matrix for spin $i$, $\rho_i = \tr_{l\ne i}(\rho^{(N)})$, is given by 
\begin{equation}
    |\rho_i(t)\rrangle = \overline{ |\rho_i(t)\rrangle_\bxi  \prod_{l\ne i} \llangle I_l|\rho_l(t)\rrangle_\bxi}
\end{equation}
Note that 
$|\rho_i\rrangle \ne \overline{|\rho_i\rrangle_\bxi}$, that is, the average over different realizations still involves all the spins. One can find similar expressions for any subset of spins.

Similar considerations about the quantum stochastic evolution of a single spin also apply to our treatment here: the stochastic quantum evolution is sign-problem free but at the expense of sampling over unphysical trajectories, yet the  density matrix remains hermitian for each trajectory; additionally, stochastic solutions exist at all times. 
In practice, the efficiency of our stochastic approach depends on the number of trajectories required for convergence.
In the next subsection, we present numerical simulations  showcasing the utility of our approach when an exact numerical computation is unavailable.

\subsection{Numerical results}
In this section, we consider the dynamics of the Dicke model where all spins are coupled to a single ``cavity mode'' with the same coupling $g_i=g$; a uniform coupling is not necessary in our approach but allows a comparison against the exact numerical simulation. 

\begin{figure}[t]
    \centering
    \includegraphics[scale=0.45]{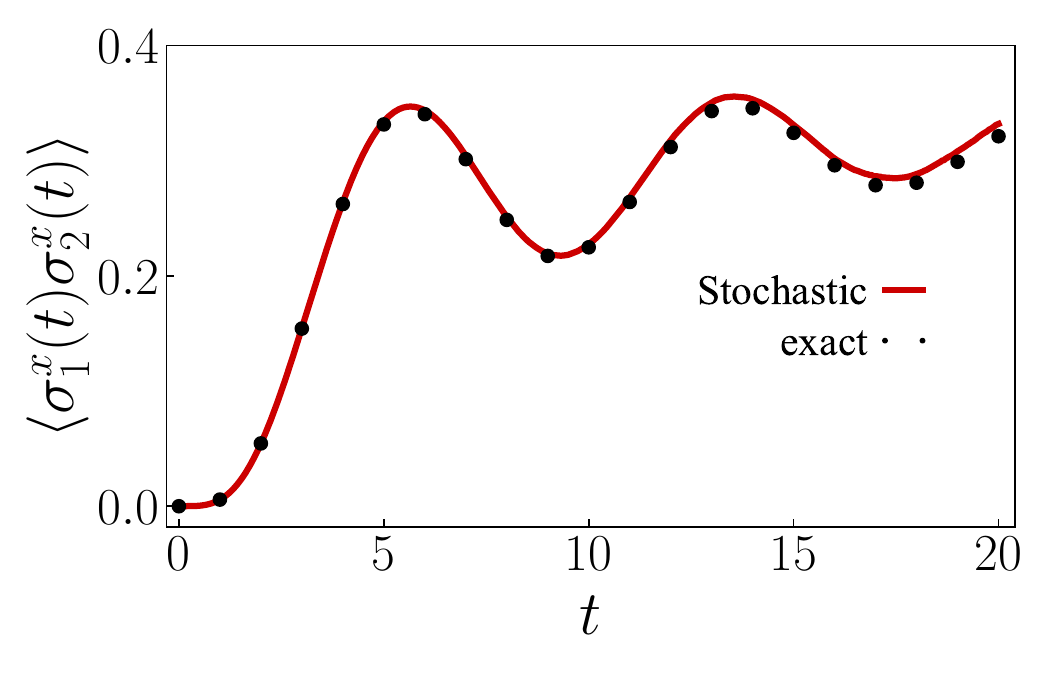}
    \caption{Stochastic evolution of the correlator $\langle \sigma^x_1 \sigma^x_2\rangle$ for the Dicke model with $N=3$ spins starting from an initial state where the spins are fully polarized along the $z$ direction and the cavity modes is in a vacuum state; here, $\omega,\kappa=1, \Delta=0.4$ and $g=0.3$.  The stochastic average of $10^8$ trajectories is well in agreement with an exact numerical computation up to $t=20$. Slight deviation at longer times is likely due to the moderate time step $dt=0.01$.}
    \label{fig:N=3}
\end{figure}

We first consider $N=3$ spins and take the parameters $\omega=\kappa=1, \Delta=0.4, g=0.3$. In \cref{fig:N=3}, we plot the two-point correlation function $\langle \sigma^x_1\sigma^x_2\rangle$ as a function of time up to $t=20$ starting from an initial state where the spins are along the positive $z$ direction (no correlations at $t=0$) and the cavity mode is in its vacuum state.
The exact dynamics is well captured by the stochastic average over $10^8$ trajectories; the slight deviation at long times is likely due to the time step $dt=0.01$. As time evolves, nontrivial correlations are formed between the spins (while $\langle\sigma^x_1 \rangle=\langle\sigma^x_2 \rangle=0$), yet the decoupled stochastic dynamics evolves the spins in a factorized form. 

\begin{figure}[t]
    \centering
    \includegraphics[scale=0.45]{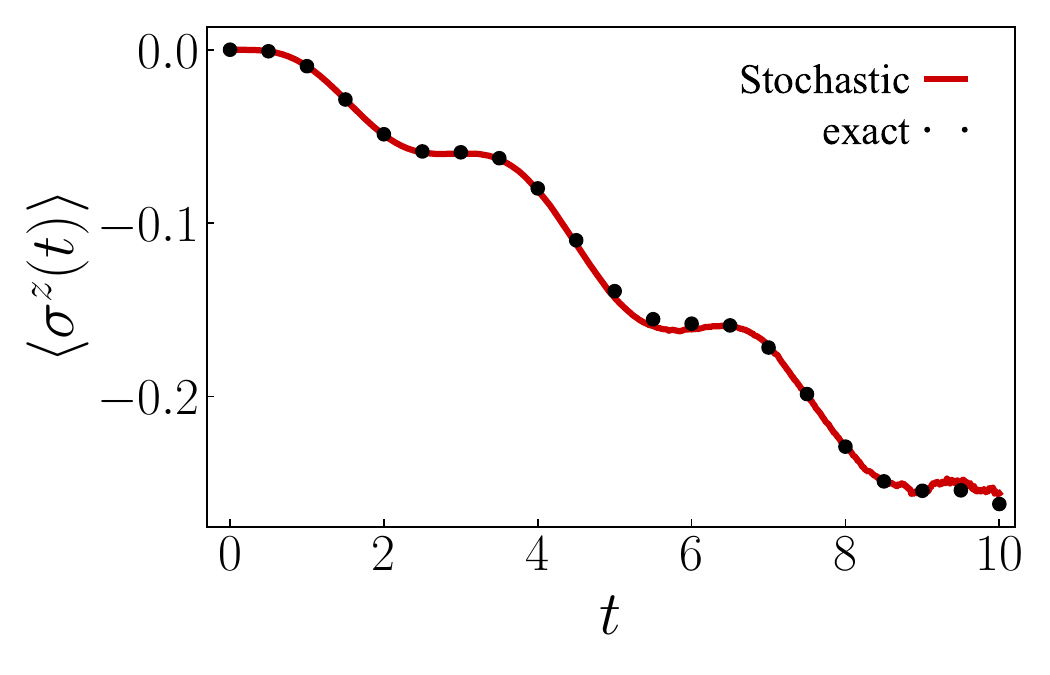}
    \caption{Stochastic evolution of $\langle\sigma^z_1(t)\rangle$ for the Dicke model with $N=30$ spins starting from an initial state where spins are fully polarized along the $x$ direction and the cavity mode is in its vacuum state; here, $\omega=\kappa=\Delta=1$ and $g=0.4$. The stochastic average of $10^8$ trajectories is  in good agreement with an exact numerical computation using the permutation symmetry of the Dicke model. The convergence at longer time can be improved by averaging over more trajectories.}
    \label{fig:N=30}
\end{figure}

As a second example, we consider the dynamics of a system of size $N=30$ starting from an initial state where the spins are fully polarized along the positive $x$ direction and the cavity mode is in its vacuum state. We take the parameters $\omega=\kappa=\Delta=1, g=0.4$ and choose a time step of $dt=0.005$. In \cref{fig:N=30}, we can simulate the dynamics up to $t\lesssim 10$ by averaging over a moderate number of trajectories ($10^8$) and find an excellent agreement with the exact result; the latter is obtained by taking advantage of the permutation symmetry and working in the Dicke manifold \cite{Kirton_review}. 
Without the permutation symmetry, a system size of $N=30$  is well outside the domain of exact diagonalization, also taking into account  the coupling to the bosonic mode and the open quantum system dynamics. 
Our method thus provides a worthwhile alternative when there is no such symmetry. 
Convergence with number of trajectories worsen as the total time or the coupling to the bosonic mode increase. Higher-order techniques for solving stochastic equations and further numerical optimizations should improve the convergence.


\section{Spin coupled to ($\infty-$)many bosons}\label{sec:N=1_M>1}
In this section, we consider a single spin coupled to several, many or possibly infinitely many, bosonic modes. We quote the Hamiltonian for completeness:
\begin{equation}\label{eq:hamiltonian_1_M}
    H_{1,M}=  \frac{\Delta}{2}  \sigma^z+\sum_{\alpha=1}^M\omega_\alpha a^\dagger_\alpha a_\alpha  +  \sigma^x \sum_{\alpha}\frac{g_{\alpha }}{2}  (a_\alpha+a_\alpha^\dagger) 
\end{equation}
Again, we assume that each bosonic mode is subject to Markovian loss characterized by the Lindblad operator $L_\alpha= \sqrt{\kappa_\alpha}a_\alpha$.
Analogously to \cref{sec:N=M=1}, we trade in the bosonic operator for stochastic fields, only now we must include  $M$ such variables to represent all the bosonic modes. 
The full stochastic evolution is then given by
\begin{equation}
    \frac{d}{dt} |\rho_S\rrangle_{\{\xi_\alpha\}} = \mathbb K^{\rm I}(t) |\rho\rrangle_{\{\xi_\alpha\}}
\end{equation}
where 
\begin{equation}\label{eq:stoch-L3}
    \mathbb K^{\rm I}(t)=\mathbb L_S +  \mathbb S\sum_\alpha ig_\alpha (\underline\psi_\alpha + \underline{\bar\psi}_\alpha)- \mathbb T\sum_\alpha \frac{g_\alpha}{2\kappa_\alpha} (\xi_\alpha+\bar\xi_\alpha) 
\end{equation}
Again, the first term on the rhs of the above equation denotes the single-spin terms. The noise variables $\xi_\alpha$ and the associated field $\underline\psi_\alpha$ are exactly determined in the same fashion as \cref{eq:langevin,eq:white_noise} with the substitution $\xi, \underline{\psi} \to \xi_\alpha, \underline{\psi}_\alpha$ and $\omega,\kappa \to \omega_\alpha, \kappa_\alpha$. To be more concrete, we have 
\begin{equation}
    \underline \psi_\alpha(t) =  i G_\alpha(t)\underline\psi_{\alpha}(0) + \int_{0}^t dt' G_\alpha(t-t') \xi(t') 
\end{equation}
where $G_\alpha(t)$ is the free (causal) Green's function corresponding to the bosonic mode $\alpha$, 
\begin{equation}
    G_\alpha(t) = \frac{1}{i\partial_t - \omega_\alpha +i\kappa_\alpha}= -i \Theta(t)e^{-i\omega_\alpha t -\kappa_\alpha t}
\end{equation}
The quantum stochastic evolution can be then solved for a given realization; a single trajectory now comprises all the noise variables $\xi_\alpha$ and the initial values for the corresponding field $\underline{\psi}_\alpha$. Finally, the physical density matrix is obtained by  averaging over many trajectories. 

While the above strategy is in principle feasible, it would be rather demanding if there are many, or even a continuum of, bosonic modes, an example of which is the paradigmatic spin-boson model where a two-level system is coupled to an infinite bath \cite{Leggett_spin_boson_model,weiss2012quantum}. 
A more efficient route is desired in this case. To this end, we first assume that the bosonic modes are initially in their vacuum state; later in this section, we extend our results to a general initial state.
Now, taking advantage of the linear  (stochastic) equation for the classical fields as well as the initial Gaussian state, one can combine all the noise variables $\xi_\alpha$ into a single variable (similarly for the associated fields $\underline{\psi}_\alpha$): 
\begin{equation}
    \underline{\Psi}(t) \equiv \sum_\alpha g_\alpha \underline\psi_\alpha,\quad \Xi(t) \equiv \sum_\alpha \frac{g_\alpha}{2\kappa_\alpha} \xi_\alpha
\end{equation}
The collective noise variable $\Xi(t)$, being a sum of white noise terms, is itself white noise with the correlations
\begin{equation}\label{eq:Xi_correlations}
    \overline{\Xi(t) \bar\Xi(t') }= \gamma \delta(t-t'), \quad \gamma= \sum_\alpha \frac{g_\alpha^2}{4\kappa_\alpha}
\end{equation}
where we have defined the dissipation rate $\gamma$. Some algebra shows that, for $t>t'$,
\begin{align}\label{eq:chi_and_C_2}
    \begin{split}
    \chi(t,t') &\equiv  \overline{\underline{\Psi}(t)\bar\Xi(t')} = \sum_\alpha \frac{g_\alpha^2}{2} G_\alpha(t-t')  \\
    C(t,t') &\equiv \overline{\underline{\Psi} (t)\bar{\underline{\Psi}}(t')}  = i\sum_\alpha \frac{g_\alpha^2}{2} G_\alpha (t-t')
    \end{split}
\end{align}
while, for $t<t'$, we have $\chi(t,t')=0$ and $C(t,t')= \bar C(t',t)$. The functions $\chi$ and $C$ define the (causal) response and correlation functions, respectively. Notice that they are both translation invariant as they only depend on $t-t'$. While this is always the case for the response function in a linear system, the correlation function becomes translation invariant as we have assumed that the bosonic modes are initially in their vacuum state. 
While $\underline{\psi}_\alpha$ is fully specified by $\xi_\alpha$ together with its initial value, the collective field $\underline{\Psi}$ is not completely determined by $\Xi$ even for  fixed initial conditions: 
this is because there are many noise realizations $\{\xi_\alpha(t)\}$ for a given $\Xi(t)$. 
This redundancy,
and consequently the uncertainty in $\underline{\Psi}$, can be encoded into another Gaussian distributed noise variable. 
More precisely, we can capture the correlators in \cref{eq:chi_and_C_2} by writing 
\begin{equation}\label{eq:Psi_from_Xi}
    \underline{\Psi}(t) = \frac{1}{\gamma}\int_0^\infty dt'\chi(t,t')\Xi(t') + X(t)
\end{equation}
where we have introduced the noise variable $X(t)$ with the correlations
\begin{align}\label{eq:checkC}
\begin{split}
    \overline{X(t) \bar X(t')} = \check C(t,t') 
\end{split}
\end{align}
while all the other (self- or cross-) noise  correlations are vanishing, $\overline{X(t) X(t')} = \overline{X(t) \Xi(t')} = \overline{X(t) \bar\Xi(t')} =0$,
and the function $\check C(t,t')$ is defined as 
\begin{equation}\label{eq:check_C}
    \check C(t,t') = C(t,t')- \frac{1}{\gamma}\int_0^\infty dt'' \chi(t-t'') \bar\chi(t'-t'')
\end{equation}
One can easily verify that \cref{eq:chi_and_C_2} follows from the definition in \cref{eq:Psi_from_Xi} and the noise correlations in \cref{eq:Xi_correlations,eq:checkC}. 
Moreover, one can show that $\check C(t,t')$ considered as a matrix (in the basis $t,t'$) is positive, in harmony with an interpretation of $X$ as Gaussian distributed colored noise (contrasted with the white noise $\Xi$); see \cref{sec:checkC}. 
If the system consists of one mode only, the function $\check C$ only captures the initial conditions which alternatively can be treated as before by sampling over the initial conditions. 
In general, $\check C(t,t')$ takes a nontrivial form and is not  translation invariant (i.e., not just a function of $t-t'$). Utilizing the hermiticity of the $C(t, t')$ matrix (in the $t, t'$ basis), we can diagonalize it as 
\begin{equation}
    \check C(t,t') = \sum_{a} \check c_a\bar \theta_a(t) \theta_a(t') 
\end{equation}
where the functions $\theta_a(t)$ define a complete basis and $\check c_a \ge 0$ denote the diagonal elements. We can then write  $X(t)=\sum_a\sqrt{\check c_a}  X_a \theta_a(t)/\sqrt{2}$ for the complex variables $X_a$ whose real and imaginary parts are drawn from a normal distribution. The average over the field $X(t)$ can be conveniently replaced by sampling $\bX= \{X_a\}$.

We can now write the full dynamics as
\begin{equation}
    \frac{d}{dt} |\rho_S\rrangle_{\Xi, \bX} = \mathbb K^{\rm I}(t) |\rho\rrangle_{\Xi, \bX}
\end{equation}
where the generator now takes the simple form
\begin{equation}
    \mathbb K^{\rm I}(t)=\mathbb L_S +i(\underline\Psi + \underline{\bar\Psi}) \mathbb S- (\Xi+\bar \Xi) \mathbb T 
\end{equation}
and the function $\underline\Psi$ is given by \cref{eq:Psi_from_Xi} together with  $X(t) = \sum_a\sqrt{\check c_a}  X_a \theta_a(t)/\sqrt{2}$. Again, we emphasize that the initial conditions are captured directly via the function $X$ and do not require additional averaging.
Finally, to obtain the physical density matrix, we should sum over different realizations.

Before closing this section, we consider an arbitrary initial state of bosons described by the joint Wigner function of all the bosonic variables,  ${\mathscr W}_0(\{\psi_{\alpha0}\})$, where $\psi_{\alpha 0}$ denotes the initial value of the field $\underline{\psi_\alpha}$. 
To account for the initial state, we should substitute  \cref{eq:Psi_from_Xi,eq:check_C} by
\begin{align}
        &\underline{\Psi}(t) = \frac{1}{\gamma}\int_0^\infty dt'\chi(t,t')\Xi(t') + X(t) + i\sum_\alpha g_\alpha G_\alpha(t) \psi_{\alpha0}  \nonumber \\
    &\overline{X(t) \bar X(t')} 
    = \check C(t,t') - \sum_\alpha \frac{g_\alpha^2}{2} G_\alpha(t)\bar G_\alpha(t') 
    \label{eq:modified_modified_C}
\end{align}
In the last equation, we have removed the contribution of the initial state to the modified correlations $\check C(t,t')$, and rather capture it directly in the definition of $\Psi(t)$ which now depends explicitly on the initial values $\{\psi_{\alpha0}\}$ which in turn are sampled according to the Wigner distribution function. One can also show that the rhs of the second line of \cref{eq:modified_modified_C} is positive as a matrix; see \cref{sec:checkC}.
For a Gaussian initial state (e.g., when bosons are initially in their vacuum state), one can absorb the Gaussian fluctuations due to the initial state in colored noise correlations as before; however, the above equations allow us to consider a general initial state. In the end, the sampling over the original $M$ white noise variables $\{\xi_\alpha(t)\}$ 
is reduced to that of a single white noise $\Xi(t)$ plus sampling $M$ initial values $\{\psi_{\alpha0}\}$. 

\subsection{Spectral function vs Markovian dissipation}
For a continuum of bosonic modes, it is convenient to introduce the  spectral function of the bath. Let us first assume no Markovian dissipation, i.e., $\kappa_\alpha=0$. 
The bath can be characterized by the spectral function defined as $J(\omega)=\pi\sum_\alpha g_\alpha^2\delta(\omega-\omega_\alpha)$. In the continuum limit, the sum over modes $\alpha$ becomes an integral. The behavior of the spectral function, especially at low frequencies,  determines the nature of the quantum bath \cite{Leggett_spin_boson_model,weiss2012quantum}. We shall consider an Ohmic bath later in this section. 

While Markovian dissipation is not typically considered in the discussion of spin-boson models \cite{Leggett_spin_boson_model,weiss2012quantum}, quantum simulation of these models often come with Markovian dissipation \cite{Porras_2008,Lemmer_2018,Burger_2022,frisk2019ultrastrong}, which is the focus of this work. We further emphasize that such models are inherently driven-dissipative as they involve both a coherent drive and incoherent loss. 
Now turning on Markovian dissipation, we  can still define the spectral function $J(\omega)$ as above, but we must incorporate the Markovian dissipation in the correlation and response functions. 
Let us first denote by $\kappa(\omega)$ the dissipation rate corresponding to a mode with the natural frequency $\omega$ (assuming that there is at most a single mode corresponding to a given frequency).
Next, we can write the constant $\gamma$ as well as the functions $\chi(t-t')\equiv\chi(t,t')$ and $C(t-t')\equiv C(t,t')$ in terms of the spectral function $J(\omega)$ and the function $\kappa(\omega)$ via the substitution $\sum_\alpha \to \int \frac{d\omega}{\pi}$, $\kappa_\alpha \to \kappa(\omega)$ and $g_\alpha^2 \to J(\omega)$:
\begin{align}
    \begin{split}
        \gamma &= \frac{1}{4}\int \frac{d\omega}{\pi}  \frac{J(\omega)}{\kappa(\omega)} \\
    \chi(t) &=  -  \frac{i}{2}\Theta(t) \int \frac{d\omega}{\pi} J(\omega) e^{-i\omega t -\kappa(\omega) t}
    \end{split}
\end{align}
We also define $C(t)= \bar C(-t) = i\chi(t)$ for $t>0$. The spectral function $J(\omega)$ together with $\kappa(\omega)$ uniquely define the spin-boson model that is further subject to Markovian loss. 

\subsection{Numerical results: Lossy Ohmic bath}
In this section, we consider the paradigmatic spin-boson model where a spin is coupled to an Ohmic bath whose spectral function \cite{Leggett_spin_boson_model,weiss2012quantum}
is given by 
\begin{align} \label{eq:eq0a1}
J(\omega)=2\pi\alpha \omega e^{-\omega/\omega_c},
\end{align}
Here, the parameter $\alpha$ defines the strength of the 
\begin{figure}[t]
    \centering
    \includegraphics[scale=0.45]{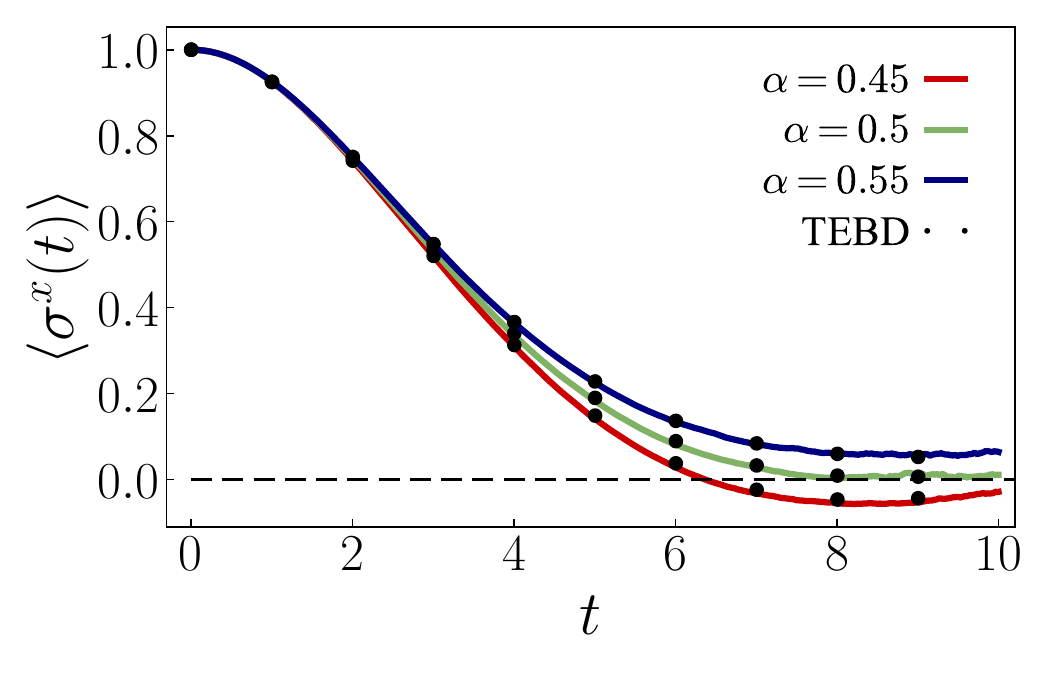}
    \caption{Stochastic evolution of $\langle \sigma^x(t) \rangle$ for a spin coupled to a lossy Ohmic bath starting from an initial state where the spin is fully polarized along the $x$ direction; here,  $r=1/2$ characterizes the ratio of loss to frequency of the bosons in the bath, $\omega_c=1$ is the bath cutoff, $\Delta =0.4$,  and $\alpha=0.45, 0.5, 0.55$ is the spin-bath coupling strength (see the text for details). The numerical results are in agreement with the TEBD simulation.}
    \label{fig:ohmic}
\end{figure}
spin-boson coupling, and $\omega_c$ defines a soft cutoff for the bath. To fully characterize the bath, we must specify the Markovian dissipation as well. To this end, we define a one-parameter family of baths characterized by the function $\kappa(\omega) = r \omega$ where the  dissipation rate for a given mode is proportional to its frequency with the constant of proportionality $r>0$.  A forthcoming paper will be dedicated to this model and studying its rich behavior \cite{kamar_dissipative_spin_boson}; here, we just take it as a testbed for our stochastic method. The bath introduced here is identified by 
\begin{align}
    \gamma = \frac{\alpha \omega_c}{2r} 
\end{align}
and the response function 
\begin{equation}
    \chi(t) =  \frac{-i\alpha\Theta(t)}{\big[\omega_c^{-1}+ (i+r)t\big]^2}
\end{equation}
Note specifically that this function smoothly interpolates to the standard Ohmic bath as $r\to0$. 

Here, we choose $r=1/2$ while considering the coupling strengths $\alpha= 0.45, 0.5, 0.55$ and   $5\times10^7, 10^8, 2 \times 10^8$ trajectories, respectively. We take the parameters $\Delta=0.4, \omega_c=1$ and choose the time step $dt=0.01$. We further consider an initial state where the spin is fully polarized along the $x$ direction and all the bosonic modes are initially in their vacuum state. The numerical results using the stochastic equation are shown in \cref{fig:ohmic} and are in very good agreement with a time-evolving block decimation (TEBD) simulation \cite{Vidal_TEBD}. For the TEBD simulation, we use the semi-infinite mapping of the spin-boson model \cite{Chin_2010} combined with a vectorization scheme of the density matrix \cite{Vidal_Mixed_State_MPS}.
Within the time window considered in the stochastic simulation, the dynamics for  $\alpha<0.5$ appears to be qualitatively different from $\alpha> 0.5$. Indeed, the standard spin-boson model (with no Markovian dissipation) is known to exhibit a transition from underdamped to overdamped dynamics exactly at $\alpha=0.5$ when $\Delta \ll \omega_c$  \cite{Leggett_spin_boson_model,Peter_Stochastic_Schrodinger_Equation,Alex_chin_TDVP_OBB,kamar_spin_boson_model_stochastic_SCH,kamar_splitting_of_Hilbert_Space}. 
Whether or not the lossy model considered here exhibits a similar transition is beyond the scope of this work, as it requires a larger number of trajectories. 
However, our stochastic approach, despite its simplicity, allows for an exact simulation at intermediate coupling strengths and time scales.
In contrast, a numerical simulation of the spin-boson model at moderate coupling strengths typically requires sophisticated computational methods such as NRG \cite{Ralf_Bulla_NRG_Spin_Boson_Model}, MPS-based methods \cite{Brockt_Optimal_Bosonic_Basis,Stolpp_OBB,Alex_chin_TDVP_OBB,kamar_splitting_of_Hilbert_Space}, among others \cite{Xu_2022}.

\section{$N$ spins coupled to $M$ bosons}\label{sec:N>1_M>1}

Finally, in this section, we consider a general spin-boson model where $N$ spins are coupled to $M$ bosons, described by the Hamiltonian in \cref{eq:hamiltonian} which we quote for completeness:
\begin{equation}\nonumber 
    H_{N,M}= \frac{\Delta}{2} \sum_{i=1}^N \sigma^z_i +\sum_{\alpha=1}^M\omega_\alpha a^\dagger_\alpha a_\alpha +  \frac{1}{\sqrt{N}}\sum_{\alpha i}\frac{g_{\alpha i}}{2} \sigma^x_i(a_\alpha+a_\alpha^\dagger)
\end{equation}
Again, we assume that each bosonic mode is subject to Markovian loss of bosons at the rate $\kappa_\alpha$.

In order to treat the many-body problem, we combine the approaches presented in \cref{sec:N>1_M=1,sec:N=1_M>1}:
\begin{enumerate}[wide, labelwidth=!, labelindent=0pt, leftmargin=0pt]
    \item First, we consider the bosonic modes separately. For a given mode $\alpha$, we follow the steps in \cref{sec:N>1_M=1} to eliminate it in terms of the classical field $\underline\psi_\alpha$ and white noise variables $\xi_\alpha$, which satisfy 
    \begin{equation}
    \begin{split}
         G_\alpha^{-1} \underline\psi_\alpha &= \xi_\alpha(t) \\ 
        \overline{ \xi_\alpha(t) \bar\xi_\alpha(t')} &= \kappa_\alpha \delta(t-t'),
    \end{split}
    \end{equation}
    with $G_\alpha$ the corresponding (causal) Green's function. Again, this generates an effective coupling between all the spins, and would be demanding for a large spin system. 
    To decouple the spins, we adopt the trick in \cref{sec:decoupling}: we introduce $N$ fictitious noise fields $\xi_{\alpha i}$ (per mode $\alpha$) such that 
    \begin{equation}
        \begin{split}
             G_\alpha^{-1} \underline\psi_\alpha &= \frac{1}{\sqrt{N}}\sum_{i=1}^N\xi_{\alpha i}(t)     \\
            \overline{\xi_{\alpha i}(t) \bar\xi_{\beta j}(t')} &= \kappa_\alpha\delta_{\alpha\beta}\delta_{ij}\delta(t-t'), 
        \end{split}
    \end{equation}
    The dynamics is then factorized and each spin is evolved as (in the It\^{o} convention) 
    \begin{align}
        \frac{d}{dt} |\rho_i\rrangle_{\{\xi_{\alpha i}\}} = \mathbb K^{\rm I}_i (t) |\rho_i\rrangle_{\{\xi_{\alpha i}\}} 
    \end{align}
    where 
    \begin{align}
    \mathbb K^{\rm I}_i (t) = \mathbb L_i +  
     \sum_\alpha\frac{ig_{\alpha i}}{\sqrt{N}}(\underline\psi_\alpha + \underline{\bar\psi}_\alpha)\mathbb S_i- \frac{g_{\alpha i}}{2\kappa_\alpha} (\xi_{\alpha i}+\bar \xi_{\alpha i}) \mathbb T_i 
    \end{align}
    The full density matrix is then obtained by averaging over different realizations of white noise variables $\{\xi_{\alpha i}(t)\}$ and the initial values $\{\underline\psi_\alpha(0)$\}. For simplicity, we assume that the initial state is a factorized state of spins and bosons, and moreover the bosons are initially in their vacuum state; a generalization to a more general initial state is straightforward and follows the prescription in the previous sections. 

    \item In the above equations, each spin is still coupled to $M$ noise variables. Following the steps in \cref{sec:N=1_M>1}, 
    we can make another transformation where the $M\times N$ noise variables $\xi_{\alpha i}$ are reduced to just $N$ variables, one for each spin. 
    To this end, we first define
    \begin{align}
        \begin{split}
            \underline\Psi_i(t) &= \frac{1}{\sqrt{N}}\sum_\alpha g_{\alpha i} \underline \psi_\alpha(t),\\
            \Xi_i(t) &= \sum_\alpha \frac{g_{\alpha i}}{2\kappa_\alpha}\xi_{\alpha i}(t)    
        \end{split}
    \end{align}
    The generator $\mathbb K^{\rm I}$ now takes a simple form as
    \begin{align}
        \mathbb K^{\rm I}_i (t) = \mathbb L_i +  
        i (\underline\Psi_i+\underline{\bar \Psi}_i) \mathbb S_i- (\Xi_i+\bar \Xi_i) \mathbb T_i 
    \end{align}
    where $\Xi_i(t)$ are the new white variables defined by the correlations
    \begin{equation}
        \overline{ \Xi_{i}(t) \bar\Xi_{j}(t')} =\gamma_i \delta(t-t'),\quad \gamma_i \equiv \sum_\alpha \frac{g_{\alpha i}^2}{4\kappa_\alpha}
    \end{equation}
    Furthermore, the function $\underline{\Psi}_i$ is now determined by 
    \begin{align}
        \underline{\Psi}_i = \frac{1}{N} \sum_j \int dt' \chi_{ij}(t,t') \Xi_j(t')/\gamma_j + X_i (t)    
    \end{align}
    where the variable $X_i(t)$ is a Gaussian-distributed random variable with the correlations 
    \begin{equation}
    \overline{ X_i(t) \bar X_j(t')} = \frac{1}{N} \check C _{ij}(t,t')
    \end{equation}
    hence, colored noise. 
    Assuming that initially the bosonic modes are in their vacuum state, the kernels ${\chi_{ij}(t,t')=\chi(t-t')}$, ${C_{ij}(t,t') =C_{ij}(t-t')}$ and $\check C_{ij}(t,t')$ are given by
    \begin{align}
        \chi_{ij}(t) &= \frac{1}{2}\sum_\alpha g_{\alpha i} g_{\alpha j} G_\alpha(t)  \\
        C_{ij}(t) &= \bar C_{ij}(-t) = i\chi_{ij}(t), \qquad t>0 \nonumber\\
        \check C_{ij}(t,t') &=  C_{ij}(t) -\frac{1}{N}\sum_l \frac{1}{\gamma_l}\int_{t''} \chi_{il} (t-t'') \bar \chi_{jl} (t'-t'') \nonumber
    \end{align}
    Again, one can show that $\check C_{ij}(t,t') \ge 0$ as matrix defined with the rows and columns defined as $it$ and $jt'$, respectively. We can then proceed as before to decompose $\check C_{ij}(t,t')$ in terms of a complete basis and write $X_i$ as a sum of different terms each with a complex-valued coefficient that is drawn from a normal distribution. 
    The quantities of interest such as expectation values of local operators or correlations function can be computed by first evolving each spin for a given realization, and then averaging over white noise $\{\Xi_i(t)\}$ as well as colored noise $\{X_i(t)\}$. Finally, a generalization to an arbitrary initial state follows analogously to the previous sections. 
\end{enumerate}

\section{Summary and outlook}\label{sec:outlook}
In this paper, we have considered a generalized spin-boson model and developed a hybrid stochastic quantum-classical approach to the evolution of the spins. 
To this end, we have traded off the bosonic modes for classical stochastic variables, which are then used as an input 
for a stochastic quantum evolution of the spins. 
In this approach, the spins are effectively decoupled for each stochastic realization, but   
this comes at the expense of sampling over unphysical states. Specifically, the density matrix is not trace 1 or positive, a fact that could hinder convergence at late times. 
However, we crucially utilize Markovian dissipation to treat the response function in a causal fashion, and to preserve hermiticity of the density matrix, a convenient feature for numerical simulations.  Our work thus provides an example where Markovian dissipation can be used as a computational resource for simulating quantum systems \cite{propp2023decoherence}. 
We have showcased the utility of our approach in scenarios where an exact numerical computation is not available. 
Our work is relevant to emerging quantum simulation platforms including trapped ions \cite{Sorensen_1999,Porras_2004,Porras_2008,Kim_2009,britton2012engineered}, cavities via cold atoms \cite{Leroux_2010,Ritsch_2013,Vaidya_2018}, superconducting qubits \cite{niemczyk2010circuit,LEHUR2016808}, and optomechanics \cite{Aspelmeyer_2014}.

Our approach extends beyond the existing methods rooted in the Feynman-Vernon influence functional and their stochastic variants \cite{Peter_Stochastic_Schrodinger_Equation,LeHur_2010,lesovik2002dynamics,LeHur_2014,kamar_spin_boson_model_stochastic_SCH,Stockburger1998,stockburger1999stochastic,Stockburger_2002,STOCKBURGER_2004,Koch_2008,LeHur_2016}, collectively referred to as the FV approach. 
First, as remarked above, our approach keeps causality, and thus hermiticity, manifest. Second, we do not place a restriction on the initial state while the FV approach generically assumes a factorized initial state and an initially thermal state of bosons. Third, the FV approach gives rise to non-local kernels (which should be sampled by colored noise) while our hybrid quantum-classical approach remains Markovian (unless we lump bosonic variables into one or several variables; see \cref{sec:N=1_M>1,sec:N>1_M>1}). Fourth, while the FV approach is typically used in scenarios where the coupling is diagonal in a given basis (e.g., the Rabi model with $H_{SB} \sim g \sigma^x (a+a^\dagger)$), our approach can be easily generalized to non-diagonal coupling (e.g., the Jaynes-Cumming model with $H_{SB}\sim \sigma^+ a +h.c.$). Fifth, our approach in principle works even when the bosonic modes are intrinsically nonlinear in which case the bosonic modes cannot be integrated out via a Gaussian integral and the FV approach is no longer applicable. As an example, we may consider dephasing for bosons, $L= \sqrt{\kappa_{dph}} a^\dagger a$, which would lead to nonlinear dynamics at the level of the Liouvillian; such dephasing can be mimicked by adding a term to $H_B$,
\begin{equation}
    H_B \to H_B + k(t) a^\dagger a
\end{equation}
where $k$ is (real-valued) white noise with the correlations $\overline {k(t)k(t')}= \kappa_{dph} \delta(t-t')$. One can then carry out the stochastic analysis presented in \cref{sec:N=M=1} to obtain a modified (It\^{o}) stochastic equation for $\underline\psi$:
\begin{equation}
    [i\partial_t -\omega -  k(t) +i \kappa] \underline\psi    = \xi(t)
\end{equation}
involving both additive [$\xi(t)$] and multiplicative [$k(t)$] white noise. To obtain the physical density matrix, we must average over both noise variables. 

Spin-boson models are particularly challenging at strong coupling where the dynamics is highly nonlinear and the system may even undergo a phase transition. Indeed, the paradigmatic spin-boson model (a spin coupled to an infinite bath) exhibits a localization phase transition \cite{Leggett_spin_boson_model,weiss2012quantum} while the Rabi/Dicke models (one/many spins coupled to a cavity mode) undergo a superradiant phase transition at strong coupling \cite{Lieb73,Hepp73}. 
It is desirable to apply our stochastic approach to study such regimes; 
however, sampling over unphysical states could lead to poor convergence with the number of trajectories at strong coupling. Our approach provides an immediate advantage by ensuring hermiticity of the density matrix which strongly constrains single realizations. One can further improve the convergence by making the stochastic equation trace preserving
\cite{Strunz_1999,Stockburger_2002,STOCKBURGER_2004}; however, the resulting nonlinear stochastic equation could lead to unstable solutions although different routes are proposed to alleviate this behavior  \cite{Koch_2008}. 
In principle, our approach only requires averaging over well-defined stochastic solutions corresponding to many trajectories which can be parallelized on classical machines; this should be contrasted with the dynamics in the full Hilbert space where an exponentially large space is required even to store the quantum state of the system. Nevertheless, the usefulness of our approach in the challenging regime of strong coupling requires a careful analysis of the scaling of the number of trajectories 
with the coupling strength along with Markovian dissipation, 
a direction that constitutes an important avenue  for future research.

\acknowledgements{}
We thank Aash Clerk for useful discussions. This work is supported by the Air Force Office of Scientific Research (AFOSR) under the award number FA9550-20-1-0073. We also acknowledge support from the National Science Foundation under the NSF CAREER Award (DMR2142866) as well as the NSF grant  PHY2112893.

\appendix

\section{Derivation of \cref{eq:q2c_mapping}}\label{sec:derivation_rho}
In this section, we derive \cref{eq:q2c_mapping}
for a single spin coupled to a cavity mode. The generalization of our approach to many spins is straightforward. 
We follow a hybrid approach to the functional integral: we first apply the quantum-to-classical mapping utilized in Refs.~\cite{Paz_2021_exact,Paz_2021_EPL} to the spin operators and map them to classical discrete variables,
and then use the phase-space approach in Ref.~\cite{foss-feig_emergent_2017} to turn the cavity operators to {$c$ numbers} using the Weyl ordering \cite{POLKOVNIKOV20101790}. 
To this end, we break the Liouvillian as 
\begin{equation}
    {\cal L} = {\cal L}_{S} + {\cal L}_{B}+ {\cal L}_{SB}
\end{equation}
with the first two terms on the rhs including only spin and bosonic terms, respectively. The last term denotes the interaction between the two and is given by 
\begin{equation}
    {\cal L}_{SB}= -i [H_{SB}, \bullet], \qquad     H_{SB}= \frac{g}{2} \sigma^x (a+ a^\dagger) 
\end{equation}
Different interactions (e.g., Jaynes-Cumming model) can be treated in a similar fashion. 
In the above equation and throughout this Appendix, we denote the action of a superoperator (such as $\cal L$) on an arbitrary operator by the location and ordering of the symbol $\bullet$.
Next, we Trotterize the evolution as 
\begin{equation}\label{eq:trotter}
    \rho(t) = e^{t{\cal L} }(\rho_0) = \underbrace{e^{\delta t \cal L}(e^{\delta t \cal L}(e^{\delta t \cal L}(\cdots (e^{\delta t \cal L}}_{n \,\, \mbox{\scriptsize times}}(\rho_0))\cdots)
\end{equation} 
with $\rho_0$ the initial state at time $t=0$ and $n =t/\delta t$. 
Note that the Liouvillian is a superoperator, and the above expression is not a matrix multiplication. 
To carry out the quantum-to-classical mapping for the spin, we 
we must insert a complete basis for spin operators at each time slice. We introduce the identity superoperator 
\begin{equation}\label{eq:id_superoperator}
    {\cal I}_S = \sum_{\sigma^u = \pm 1}|\sigma^u \rangle\langle \sigma^u|\bullet \sum_{\sigma^l = \pm 1} |\sigma^l \rangle\langle \sigma^l|
\end{equation}
where $\sigma^x |\sigma\rangle = \sigma|\sigma\rangle$ with $\sigma$ representing either $\sigma^u$ or $\sigma^l$; the notation for the superscript is inspired by the upper/lower branches of the Keldysh contour. 
Inserting the identity superoperator at each time slice in \cref{eq:trotter}, we have 
\begin{align}
        \langle \sigma| \rho(t)|\sigma'\rangle 
    = &\sum_{\boldsymbol \sigma} \Bigg\{\big[\prod_{k=0}^{n-1}  \langle \sigma_{k+1}^u |\left(e^{\delta t{\cal L}} 
    (|\sigma^{u}_k\rangle\langle \sigma^l_k|) \right)|\sigma^{l}_{k+1}\rangle\big] \nonumber\\
    &\times  \langle \sigma^u_0 |\rho_0 |\sigma^l_0\rangle \Bigg\}
\end{align}
Here, we have introduced the complete basis $\sigma_k^{u/l}$ at the Trotter time $k$, and
summed over all ${\boldsymbol\sigma}= (\sigma^{u/l}_0, \sigma^{u/l}_1, \cdots, \sigma^{u/l}_{n-1})$, and further identified $\sigma^{u/l}_n =\sigma, \sigma'$, respectively. We emphasize that $\langle \sigma'| \rho(t)|\sigma\rangle $ is still a matrix in the Hilbert space of the bosonic operator.
At this point, we carry out the standard Trotter-Suzuki expansion and write 
\begin{equation}
    e^{\delta t{\cal  L}} = e^{\delta t ({\cal L}_{B}+{\cal L}_{S})} e^{\delta t {\cal L}_{SB}}+ {\cal O}(\delta t^2)
\end{equation}
The treatment of the spin-boson coupling is now straightforward within the quantum-to-classical mapping since this coupling is diagonal in the basis adopted in \cref{eq:id_superoperator}: 
\begin{equation}
    e^{\delta t{\cal L}_{SB}}(|\sigma^{u}_k\rangle\langle \sigma^l_k|) = |\sigma^{u}_k\rangle\langle \sigma^l_k| e^{-i\delta t g \sigma^u_k (a+a^\dagger)\bullet + i\delta t g\sigma^l_k  \bullet (a+a^\dagger)} 
\end{equation}
where the notation in the exponent should be understood as follows: defining the  exponential in a series expansion, we find a nested product of (bosonic) superoperators, and the location of $\bullet$ guides the action of subsequent superoperators on an input operator. 
We can then write
\begin{widetext}
    \begin{align}\label{eq:rho_mat}
        \langle \sigma| \rho(t)|\sigma'\rangle 
    = \sum_{\boldsymbol \sigma} \left\{\Big[\prod_{k=0}^{n-1}  \langle \sigma_{k+1}^u |\left(e^{\delta t{\cal L}_{S}} 
    (|\sigma^{u}_k\rangle\langle \sigma^l_k|) \right)|\sigma^{l}_{k+1}\rangle\Big]   \langle \sigma^u_0 |\rho_{S}(0) |\sigma^l_0\rangle  \times {\rm T}_k \prod_{k=0}^{n-1} e^{\delta t \widetilde{\cal L}_{B}(t_k)}(\rho_{B}(0))\right\}
    \end{align}    
\end{widetext}
where ${\rm T}_k$ is the time-ordering operator (in discretized time indexed by $k$), and we have defined the `tilde' superoperator
\begin{equation}\label{eq: L'}
    \widetilde{\cal L}_{B}(t_k) = {\cal L}_{B} -\frac{i g}{2}\left[ \sigma^u_k (a+a^\dagger)\bullet - \sigma^l_k  \bullet (a+a^\dagger)\right]
\end{equation}
with $t_k = k \delta t$.  We have additionally assumed that the initial state is factorized, $\rho_0 = \rho_{S}(0)\otimes  \rho_{B}(0)$. 
At this stage, $\sigma^{u/l}_k$ are just numbers and their ordering in the above expression is unimportant. Notice that the first product on the rhs of \cref{eq:rho_mat} only involves the spin variables, while the information about the bosonic mode as well as the spin-boson coupling is included in the second product that involves the Liouvillian $\widetilde{\cal L}_B(t_k)$. 

Next, we map the bosonic operators into classical variables in order to construct the functional integral. Specifically, we utilize a quantum-optics based approach by working in the Weyl representation where the state is represented in terms of the Wigner function (in contrast with the Keldysh functional integral where the operators are replaced by $Q$ symbols). The advantage of this method is that an unambiguous
continuous-time limit emerges naturally from a properly defined (i.e., discretized) functional integral. To this end, we follow \cite{foss-feig_emergent_2017} to establish the mapping to a functional integral for the bosonic mode. 
We shall not repeat these steps here, and just quote the resulting functional integral; an interested reader is referred to Appendix A of Ref.~\cite{foss-feig_emergent_2017}. 
In this mapping, we map bosonic operators to phase space variables $\phi_k, \psi_k$ and  the superoperator to a classical action. The evolution of the bosonic mode can be  written as a functional integral over the phase-space variables weighted by the exponential of the corresponding action (similar to Feynman path integral). 
The Liouvillian ${\cal L}_B$ describing free bosonic mode is then mapped to the classical action
 \begin{equation}
     {\mathscr S}_B= \sum_{k=0}^{n-1} \delta t {\mathscr L}_{B} (\psi_{k+1},\psi_k, \phi_k)
 \end{equation}
where we have defined the Lagrangian ${\mathscr L}_B$
\begin{widetext}
    \begin{align}
    {\mathscr L}_{B} (\psi_1,\psi_0, \phi_0) &= 2i \bar \phi_0 (\psi_1-\psi_0)/\delta t -2i \phi_0 (\bar \psi_1-\bar \psi_0)/\delta t-{\mathscr H}_B(\psi_0+\phi_0)+{\mathscr H}_B(\psi_0-\phi_0) \nonumber \\
    &\quad +i\kappa (2\bar \phi_0 \phi_0 -\phi_0 \bar\psi_0+\bar \phi_0\psi_0) \nonumber \\
    &=  2\bar \phi_0 [i(\psi_{1} -\psi_0) - \omega \psi_0 - \kappa \psi_0]+c.c. + 4i \kappa \bar\phi_0 \phi_0, \label{eq:1st_line}
\end{align}    
\end{widetext}
and ${\mathscr H}_B$ denotes the Weyl symbol corresponding to the Hamiltonian $H_B$. Notice that the sign difference between the two terms involving ${\mathscr H}_B$ is due to the same sign in the commutator $[H_B, \bullet]$. 
In the last equality, we have used the transformation
\(
    H_B =\omega a^\dagger a \to 
     {\mathscr H}_B(\alpha)= \omega(|\alpha|^2 -\frac{1}{2}),
\)
and have dropped the constant term.
Next, we consider the interaction, i.e., the expression in the bracket in \cref{eq: L'}. A similar treatment leads to the substitution $ (a+a^\dagger) \bullet \to \psi_k+ \phi_k+ c.c.$ and $\bullet(a+a^\dagger) \to \psi_k- \phi_k+ c.c.$ at time $t_k$ in the functional integral.
Now, the action involving both the free bosonic part as well as the spin-boson coupling takes the form 
\begin{widetext}
\begin{equation}\label{eq:S_B_Appendix}
    \widetilde{\mathscr S}_{B}= {\mathscr S}_{B} + \sum_{k=0}^{n-1} \Big[ 
    - \frac{g}{2} \sigma^u_k \, (\psi_k+\phi_k+c.c.) +\frac{g}{2}\sigma^l_k \, (\psi_k-\phi_k+c.c.) \Big]
\end{equation}
Mapping out the bosonic operator to phase-space variables, we can now write the spin reduced density matrix $\rho_S(t)=\tr_B(\rho(t))$ as
\begin{align}
    \begin{split}
     \langle \sigma| \rho_{S}(t)|\sigma'\rangle 
    &= \int d^2\psi_{n}\prod_{k=0}^{n-1} \frac{4d^2\psi_k d^2\phi_k }{\pi^2} e^{i \widetilde{\mathscr S}_{B}} {\mathscr W}_0 (\psi_0)\\
    &\times \sum_{\boldsymbol \sigma} \left\{\big[\prod_{k=0}^{n-1}  \langle \sigma_{k+1}^u |\left(e^{\delta t{\cal L}_{S}}  
    (|\sigma^{u}_k\rangle\langle \sigma^l_k|) \right)|\sigma^{l}_{k+1}\rangle\big] \times  \langle \sigma^u_0 |\rho_{S}(0) |\sigma^l_0\rangle \right\} \\
    &= \int d^2\psi_{n}\prod_{k=0}^{n-1} \frac{4d^2\psi_k d^2\phi_k }{\pi^2} e^{i {\mathscr S}_{B}} {\mathscr W}_0 (\psi_0)\\
    &\times \sum_{\boldsymbol \sigma} \left\{\big[\prod_{k=0}^{n-1}  \langle \sigma_{k+1}^u |\left(e^{\delta t \widetilde{\cal L}_{S}(t_k)}  
    (|\sigma^{u}_k\rangle\langle \sigma^l_k|) \right)|\sigma^{l}_{k+1}\rangle\big] \times  \langle \sigma^u_0 |\rho_{S}(0) |\sigma^l_0\rangle \right\}
    \end{split}
\end{align}
with ${\mathscr W}_0$ the Wigner function of the initial state; we have used the notation $d^2z =d\re(z) d\im(z)$. In the last equality, we have restored the free bosonic action, $\widetilde {\mathscr S}_{B} \to {\mathscr S}_{B}$, and absorbed the difference in  ${\cal L}_{S}\to \widetilde{\cal L}_{S}(t_k)$ where we have defined the `tilde' spin superoperator 
\begin{align}
    \widetilde {\cal L}_{S}(t_k) = {\cal L}_{S} -  \frac{ig}{2} \left[(\psi_k+  \phi_k+c.c.)\sigma^x \bullet - (\psi_k- \phi_k+c.c.) \bullet \sigma^x\right]  
\end{align}
Notice that this expression does not involve the indices $\sigma^{u/l}_k$ explicitly, but the spin operator $\sigma^x$ acting on a ket or a bra state at time $k$ (as prescribed by the ordering with respect to $\bullet$) reproduces the expression in \cref{eq:S_B_Appendix}.
We can then explicitly sum over the spin indices $\bsigma$ as they sum to an identity superoperator at all time slices before time $t$. The resulting density matrix is given by 
\begin{align}
     \langle \sigma| \rho_{S}(t)|\sigma'\rangle =
    \int  {\mathscr D}[\psi,\phi] e^{i{\mathscr S}_B} {\mathscr W}_0 (\psi_0) \langle\sigma| {\rm T}_k e^{\sum_{k=0}^{n-1} \widetilde{\cal L}_S(t_k)}(\rho_S(0))|\sigma'\rangle
\end{align}
where we have defined the shorthand 
\begin{equation}
    {\mathscr D}[\psi,\phi]=\int d^2\psi_{n}\prod_{k=0}^{n-1} \frac{4d^2\psi_k  d^2\phi_k }{\pi^2}
\end{equation}
as the proper measure of the path integral in discretized time. 
Finally, vectorizing the density matrix, we obtain 
\begin{align}
    |\rho_S(t)\rrangle = \int  {\mathscr D}[\psi,\phi] e^{i{\mathscr S}_B} {\mathscr W}_0 (\psi_0) {\rm T}_k e^{\sum_{k=0}^{n-1} \widetilde{\mathbb L}_S(t_k)}|\rho_S(0)\rrangle
\end{align}
where the matrix $\widetilde{\mathbb L}_S(t_k)$ corresponds to the superoperator $\widetilde{\cal L}_S(t_k)$ in the vectorized notation, and is explicitly given by 
\begin{equation}
    \widetilde{\mathbb L}_S(t_k)= \mathbb L_S -  \frac{ig}{2} \left[(\psi_k+  \phi_k+c.c.)\sigma^x \otimes I - (\psi_k- \phi_k+c.c.) I\otimes  \sigma^x\right]  
\end{equation}
Taking the limit of continuous time, one recovers \cref{eq:q2c_mapping} and can explicitly verify  that $\widetilde{\mathbb L}_S(t_k) \to \mathbb L_S + \mathbb L_{\rm int}(t)$ where $\mathbb L_{\rm int}$ is defined in \cref{eq:tilde_L_2}. 
\end{widetext}

\section{Derivation of \cref{eq:q2c_4}}\label{sec:derivation_rho_xi}
In this section, we provide the derivation of \cref{eq:q2c_4}.
To this end, we first use the Hubbard-Stratonovich transformation given by \cref{eq:HS} but in discretized time:
\begin{equation}\label{eq:HS_App}
    e^{-4\kappa \delta t \bar \phi_k\phi_k} = \frac{\delta t}{\kappa \pi}\int d^2\xi_k e^{-\delta t \bar \xi_k \xi_k  /\kappa- 2i\delta t  (\bar \xi_k \phi_k+ \xi_k \bar \phi_k)}
\end{equation}
where we have introduced the Gaussian distributed  noise $\xi_k$. Ultimately, we like to trade in the field $\phi_k$ for the noise $\xi_k$. But, as remarked before, the field $\phi_k$ also appears in the $\widetilde{\mathbb L}_S(t_k)$ matrix. To deal with the latter,
we use the Suzuki-Trotter expansion to write:
\begin{align}
    \begin{split}
        e^{\delta t\widetilde{\mathbb L}_S(t_k)} &= e^{\delta t \mathbb L_1(t_k) +  ig\delta t  (\phi_{k}+\bar\phi_k)\mathbb T } \\
    &= e^{\delta t \mathbb L_1(t_k) } e^{ig \delta t  (\phi_k+\bar \phi_{k})\mathbb T } +{\cal O}(\delta t^{3/2})
    \end{split}
\end{align}
where in the first equality we have defined $ \mathbb L_1(t_k) = \mathbb L_S+ g  (\psi_k + \bar \psi_k) \mathbb S$ for notational convenience. 
We note that the error is of the order of $\delta t^{3/2}$, and not $\delta t^2$ as one might naively expect. This is because  $\phi_k \sim 1/\delta t^{1/2}$ given the Gaussian fluctuations of the field $\phi_k$ in \cref{eq:HS_App}.
In fact, one should be careful when expanding the exponential to the lowest order. Specifically, expanding $\exp[i\delta t (\phi_q+\bar \phi_q) \cdots] \approx 1+ i\delta t (\phi_q+\bar \phi_q) \cdots$ captures terms up to $\delta t^{1/2}$ but misses the term of the order $\delta t$. In using the Suzuki-Trotter expansion, we have  ensured that the error is smaller than $\delta t$. 
We then have 
\begin{widetext}
    \begin{align}
   |\rho_S(t)\rrangle =  \int {\mathscr D}[\psi, \phi,\xi ]{\mathscr W}_0(\psi_0)e^{i {\mathscr S}'_{B}+\sum_{k=0}^{n-1} -\delta t \bar \xi_k \xi_k  /\kappa- 2i\delta t  (\bar \xi_k \phi_k+ \xi_k \bar \phi_k)}  {\rm T}_k \,\, \prod_{k=0}^{n-1}e^{\delta t \mathbb L_1(t_k) } e^{ig\delta t  \mathbb T (\phi_k+\bar\phi_k)} |\rho_S(0)\rrangle
\end{align}    
where ${\mathscr S}'_{B}$ refers to action for the free bosonic mode excluding  the term $4i \kappa \bar \phi \phi$ in the corresponding Lagrangian (\ref{eq:1st_line}). We have also defined the shorthand
\begin{equation}
    {\mathscr D}[\xi] = \prod_{k=0}^{n-1} \frac{\delta t}{\kappa \pi}  d^2\xi_k 
\end{equation}
as the measure of the functional integral over noise. 

Next, we  eliminate $\phi_k$ in the time-ordered product by writing it as a derivative with respect to $\xi_k$ as 
\begin{align}
   |\rho_S(t) \rrangle=  \int {\mathscr D}[\psi, \phi,\xi ] {\mathscr W}_0(\psi_0)e^{i  {\mathscr S}'_{B}-\sum_k \delta t \bar \xi_k \xi_k / \kappa}
    \,\,{\rm T}_k \,\, \prod_{k=0}^{n-1}e^{\delta t \mathbb L_1(t_k) } e^{-\frac{g}{2}\mathbb T (\partial_{\xi_k}+\partial_{\bar \xi_k})}  e^{-2i\delta t  (\bar \xi_k \phi_k + \xi_k \bar \phi_k)\mathbb I}|\rho_0 \rrangle \nonumber
\end{align}
An integration by parts with respect to the derivatives just introduced yields 
\begin{align}
    |\rho_S(t) \rrangle&=  \int {\mathscr D}[\psi, \phi,\xi ] {\mathscr W}_0(\psi_0)e^{i  {\mathscr S}'_{B} -\sum_k 2i\delta t  (\bar \xi_k \phi_k + \xi_k \bar \phi_k)}
   \,\,{\rm T}_k \prod_{k=0}^{n-1}e^{\delta t \mathbb L_1(t_k)  } e^{\frac{g}{2}\mathbb T(\partial_{\xi_k}+\partial_{\bar \xi_k})}  e^{- \delta t \bar \xi_k \xi_k  /\kappa \mathbb I } |\rho_S(0) \rrangle \nonumber\\
   &=\int {\mathscr D}[\psi, \phi,\xi ] {\mathscr W}_0(\psi_0)e^{i  {\mathscr S}'_{B} -\sum_k 2i\delta t  (\bar \xi_k \phi_k + \xi_k \bar \phi_k)} \,\, {\rm T}_k \prod_{k=0}^{n-1}e^{\delta t \mathbb L_1(t_k)  }  e^{- \delta t (\bar \xi_k \mathbb I+ \frac{g}{2}\mathbb T) (\xi_k \mathbb I+\frac{g}{2}\mathbb T)   /\kappa} |\rho_S(0) \rrangle
   \\
   &= \int {\mathscr D}[\psi, \phi,\xi ] {\mathscr W}_0(\psi_0)e^{i  {\mathscr S}'_{B} -\sum_k 2i\delta t  (\bar \xi_k \phi_k + \xi_k \bar \phi_k) - \delta t \bar \xi_k \xi_k  /\kappa}
   \,\,{\rm T}_k \,\, \prod_{k=0}^{n-1}e^{\delta t \mathbb L_1(t_k) - \delta t (g/2\kappa)  (\xi_k+\bar \xi_k) \mathbb T- \delta t(g^2/4\kappa) \mathbb T^2  } |\rho_S(0) \rrangle \nonumber
\end{align}
\end{widetext}
In the second line, we used the fact that $e^{a\partial_\xi}$ is a translation operator, i.e.,  $e^{a \partial_\xi} f(\xi)= f(\xi+ a)$ for a function $f(\xi)$.  In the third line, 
we have used the Suzuki-Trotter expansion again, this time to combine the exponents, and ignored an  error of the order $\delta t^{3/2}$. 
Finally, we can integrate over $\phi_k$ for all $k=0,1, \cdots, n-1$ to find the delta functions 
\begin{align}
\begin{split}
   &\int \frac{4d^2\phi_k}{\pi^2} e^{2i (\bar \phi_k f + \phi_k \bar f)} = \delta^2(f) \\
    &\mbox{with}\,\, f= i(\psi_{k+1}- \psi_k)+ \delta t(- \omega\psi_k +i\kappa \psi_k -\xi_k) 
\end{split}
\end{align}
where we have defined $\delta^2(z)=\delta(\re z) \delta (\im z)$. Integrating over  $\psi_k$  for $k=1, \cdots, n$ (excluding $k=0$) and  $\phi_k$ for $k=0,\cdots, n-1$, 
the field $\phi_k$ completely drops out while the field $\psi_k$ (for $k>0$) becomes constrained by the equation
\begin{equation}
    i(\underline\psi_{k+1}- \underline\psi_k)/\delta t- \omega\underline\psi_k +i\kappa \underline\psi_k=\xi_k
\end{equation}
and the initial condition $\psi_0$ that is drawn from  the Wigner function ${\mathscr W}_0(\psi_0)$.
We have denoted the solution to the above equation by $\underline\psi_k$. Together with the fact that the Jacobian is 1, we finally find 
\begin{widetext}
    \begin{align}\label{eq:rho_s_App}
   |\rho_S(t)\rrangle
   =& \int {\mathscr D[\xi ]} \int d^2\psi_0 {\mathscr W}_0(\psi_0) e^{-\sum_k \delta t \bar \xi_k \xi_k  /\kappa} {\rm T}_k \,\, \prod_{k=0}^{n-1}e^{\delta t (\mathbb L_S + ig (\underline \psi_k+ \underline{\bar\psi}_k)\mathbb S  -  (g/2\kappa)  (\xi_k+\bar \xi_k) \mathbb T - (g^2/4\kappa) \mathbb T^2) }  |\rho_S(0) \rrangle
\end{align}    
\end{widetext}
The exponent in this equation gives the kernel $\mathbb K(t_k)$ defined in \cref{eq:K-tilde} albeit at a discrete time step $k$. This completes our derivation of \cref{eq:q2c_4}.

Next, we derive the local generator of the dynamics. First, we define the time ordered product in the above expression as $ |\rho_S(t)\rrangle_\xi$ such that 
\begin{align}
    |\rho_S(t)\rrangle = \int {\mathscr D[\xi ]} \int d^2\psi_0 {\mathscr W}_0(\psi_0) e^{-\sum_k \delta t \bar \xi_k \xi_k  /\kappa}  |\rho_S(t)\rrangle_\xi
\end{align}
It follows from this definition that 
\begin{align}\label{eq:rho_k+1}
    &|\rho_S(t_{k+1})\rrangle_\xi= \\
    &e^{\delta t [\mathbb L_S + ig (\underline \psi_k+ \underline{\bar\psi}_k)\mathbb S  -  (g/2\kappa)  (\xi_k+\bar \xi_k) \mathbb T - (g^2/4\kappa) \mathbb T^2] }  |\rho_S(t_k) \rrangle_\xi \nonumber 
\end{align}
We then expand the exponential in powers of $\delta t$ and take the limit $\delta t\to 0$.
Specifically, we must expand  
the term proportional to noise ($\xi_k+\bar \xi_k$) to the second order since $\overline{\xi_k \bar \xi_k} = \kappa /\delta t$. The term generated at  this order is given by  
\begin{align}
    \frac{1}{2}\left(\frac{g}{2\kappa}\right)^2\delta t^2 (\xi_k+\bar\xi_k)^2\mathbb T^2  \xrightarrow[\delta t\to 0]{} \delta t \frac{g^2}{4\kappa}\mathbb T^2
\end{align}
The limit is obtained in the sense that $\delta t^2\overline{ \xi_k \bar\xi_k} =\delta t$ while all the higher \textit{cumulants} are proportional to  higher powers of $\delta t$. In essence, this statement is the same as the proof that a Wiener process is described by $dW^2 = dt$ and $dW^{2+N}=0$ for $N>0$ \cite{gardiner1985handbook}. Next, we note that the above expression cancels out against the last term of the exponent in \cref{eq:rho_k+1} once expanded to the linear order in $\delta t$. It follows that 
\begin{widetext}
    \begin{align}
    |\rho_S(t_{k+1})\rrangle_\xi = |\rho_S(t_{k})\rrangle_\xi + \delta t \left[\mathbb L_S+ ig (\underline \psi_k+ \underline{\bar\psi}_k)\mathbb S  -  (g/2\kappa)  (\xi_k+\bar \xi_k) \mathbb T\right]|\rho_S(t_{k})\rrangle_\xi
\end{align}    
\end{widetext}
In the continuum limit, this expression gives the dynamics with the generator $\mathbb K^{\rm I}$ defined in \cref{eq:K_S_Ito}. The discretized version of this equation makes it clear that the dynamics is given in the It\^{o} sense. 
As a final remark, it is straightforward to see that, using the transformation between the It\^{o} and the Stratonovich rules, the kernel $\mathbb K$ describes the stochastic dynamics in the Stratonovich sense.

\section{Influence functional and its variants}\label{sec:FV}
In this section, we show that our stochastic approach gives rise to the expected Feynman-Vernon influence functional when the initial state is factorized and the bosonic mode(s) are initially in their vacuum state. Furthermore, we utilize the Feynman-Vernon framework to verify our stochastic formulation in cases where the initial state state is arbitrary. 

\begin{widetext}
\subsection{The case $N=M=1$}\label{sec:FV_1}
We start by `integrating out' the noise as well as the initial state from  \cref{eq:rho_s_App}. 
We first introduce the notation (also see \cref{sec:Model})
\begin{equation}\label{eq:eta_k_App}
    \eta_k = (\sigma^u_k+\sigma^l_k)/2, \qquad  \tilde \eta_k = (\sigma^u_k- \sigma^l_k)/2
\end{equation}
Analogously to \cref{sec:derivation_rho}, we can insert a complete basis to perform a quantum-to-classical mapping, albeit in the basis defined above. The matrices $\mathbb S, \mathbb T$ become diagonal in this basis with the elements ${\llangle \eta \tilde\eta\mathbb |\mathbb S|\eta\tilde\eta\rrangle = -\tilde\eta}$ and ${\llangle \eta \tilde\eta\mathbb |\mathbb T|\eta\tilde\eta\rrangle = -\eta}$.
It is easy to verify that the functional integral in \cref{eq:rho_s_App} can be cast as  
\begin{align}
    \begin{split}
         \langle \sigma| \rho_{S}(t)|\sigma'\rangle =
     &\sum_{{\boldsymbol\eta},\boldsymbol{\tilde\eta}}\int d^2\psi_0 {\mathscr W}_0(\psi_0)\int {\mathscr D}[\xi] e^{-\sum_k \delta t \bar \xi_k \xi_k  /\kappa}  \llangle \sigma\sigma'|\eta_{n}\tilde\eta_n\rrangle  \llangle \eta_0\tilde\eta_0 |\rho_S(0)\rrangle\times  \\
    & \times \prod_{k=0}^{n-1}
     \llangle \eta_{k+1}\tilde\eta_{k+1}|e^{\delta t \mathbb L_{S}}|\eta_{k} \tilde\eta_k \rrangle 
     e^{-i\delta t  g \tilde\eta_k (\underline\psi_k+\underline{\bar\psi}_k)+\frac{g}{2\kappa}\delta t  \eta_k(\xi_k+\bar\xi_k)-\frac{g^2}{4\kappa} \delta t\eta_k^2 } 
    \end{split}
\end{align}    
where $\boldsymbol \eta = \{\eta_0, \eta_1, \cdots, \eta_{n}\}$. 
Here, we assume that the bosonic mode is initially in the vacuum state described by the Wigner function ${\mathscr W}_0(\psi_0)=\frac{2}{\pi}e^{-2|\psi_0|^2}$.
Now integrating over the noise $\{\xi_k$\} as well as the Wigner distribution function, we find 
\begin{align}
     &\langle \sigma| \rho_{S}(t)|\sigma'\rangle \nonumber
     =
     \sum_{{\boldsymbol\eta},\boldsymbol{\tilde\eta}}\llangle \sigma\sigma'|\eta_{n}\tilde\eta_n\rrangle  \llangle \eta_0\tilde\eta_0 |\rho_S(0)\rrangle\times    \\
    &\times  \prod_{k=0}^{n-1}
     \llangle \eta_{k+1}\tilde\eta_{k+1}|e^{\delta t \mathbb L_{S}}|\eta_{k} \tilde\eta_k \rrangle  e^{-\frac{g^2}{4\kappa} \delta t \eta_k^2 } e^{\sum_{k,k'} \delta t^2[-\frac{g^2}{2}(\overline{\underline\psi_k  \underline{\bar\psi}_{k'}} +c.c.) \tilde \eta_k \tilde \eta_{k'}-i\frac{g^2}{2\kappa}(\overline{\underline\psi_k \bar\xi_{k'} } +c.c.)\tilde \eta_k \eta_{k'} + \frac{1}{2}\frac{g^2}{4\kappa^2}(\overline{\xi_k\bar\xi_{k'}}+c.c.)\eta_{k} \eta_{k'}]} \nonumber \\
     &=  \sum_{{\boldsymbol\eta},\boldsymbol{\tilde\eta}}\llangle \sigma\sigma'|\eta_{n}\tilde\eta_n\rrangle  \llangle \eta_0\tilde\eta_0 |\rho_S(0)\rrangle
     \prod_{k=0}^{n-1}
     \llangle \eta_{k+1}\tilde\eta_{k+1}|e^{\delta t \mathbb L_{S}}|\eta_{k} \tilde\eta_k \rrangle e^{-\frac{g^2}{2} \int_{t,t'}\tilde \eta(t)\tilde \eta(t') {i\cal G^K}(t,t')} e^{-i g^2 \int_{t,t'}\tilde \eta(t)\eta(t'){\cal G}(t,t')}  \label{eq:Zeta}
\end{align}
where the overline in the first equality indicates the Gaussian average over both the noise $\xi_k$ as well as $\psi_0$, and the continuous time in the last line is a  shorthand for the sum over discrete time. Notice that the term proportional to $\eta_k^2$ cancels out against the contribution from the noise average.
Moreover, we have defined the Green's functions
\begin{align}
\begin{split}
    &{\cal G}(t,t')=\frac{1}{2} G(t-t')+c.c. =  -\sin[\omega (t-t')]e^{-\kappa(t-t')}\Theta(t-t') \\
    &i{\cal G}^K(t,t')=\kappa\int_{t''}G(t-t'') \bar G(t'-t'') +c.c. + \frac{1}{2}G(t)\bar G(t')+c.c. = \cos[\omega (t-t')] e^{-\kappa|t-t'|} \label{eq:line_2}
\end{split}
\end{align}

These Green functions correspond to the correlation and response function of the first quadrature of the cavity mode. While $i{\cal G}^K$ is purely real, we have included a factor of $i$ in harmony with the Keldysh convention. The above expressions are consistent with the influence functional upon identifying $g^2{\cal G} \to -L_1 $ and $ig^2{\cal G}^K \to L_2$; see \cref{eq:FV_influence_func}. Notice that the factor of $1/2$ in front of ${\cal G}^K$ in the exponent of \cref{eq:Zeta} is due to the symmetrization with respect to $t,t'$.

\subsection{The case $N>1$ and $M=1$}\label{sec:FV_2}
Next, we consider $N$ spins coupled to one bosonic mode. We shall assume a factorized initial state (an extension thereof is straightforward), but consider a general initial state for the bosonic mode that is not necessarily Gaussian. 
We can derive a similar functional integral expression similar to \cref{eq:rho_s_App} only with the matrix $\mathbb K$ identified as 
\begin{align}
    \mathbb K = \sum_{i=1}^N \left[\mathbb L_i +  i \frac{g_i}{\sqrt{N}}(\underline\psi + \underline{\bar\psi})\mathbb S_i- \frac{g_i}{2\sqrt{N}\kappa} (\xi+\bar \xi) \mathbb T_i\right] -\frac{1}{4N\kappa} \Big(\sum_ig_i\mathbb T_i\Big)^2
\end{align}
Inserting a complete basis as the last subsection and performing a quantum-to-classical mapping, we find 
\begin{align}
    \begin{split}
            \langle \vec \sigma| \rho_{S}(t)|\vec\sigma'\rangle  = 
    &\int d^2\psi_0 {\mathscr W}_0(\psi_0)\int {\mathscr D}[\xi] e^{-\sum_{k} \bar \xi_{k} \xi_{k}/\kappa}\sum_{\boldsymbol\eta,\boldsymbol{\tilde\eta}}\llangle \vec \sigma,\vec\sigma'|\prod_i |\eta_{i,n}\tilde\eta_{i,n}\rrangle  \llangle \eta_{i,0}\tilde\eta_{i,0} |\rho_i(0)\rrangle  \times \\
    &\times\Big[\prod_{i,k}\llangle \eta_{i,k+1}\tilde\eta_{i,k+1}|e^{\delta t\mathbb L_{S}}|\eta_{i,k}\tilde\eta_{i,k} \rrangle\Big]  e^{-i\delta t  \sum_i\frac{g_i}{\sqrt{N}} \tilde\eta_{i,k} (\underline\psi_k+\underline{\bar\psi}_k)+\delta t\sum_i\frac{g_i}{2\sqrt{N}\,\kappa}  \eta_{i,k}(\xi_{k}+\bar\xi_{k})-\frac{\delta t}{4N\kappa} (\sum_i g_i\eta_{i,k})^2 } 
    \end{split}
\end{align}
where 
we have defined $\vec \sigma= \{\sigma_1, \sigma_2, \cdots, \sigma_N\}$ and $\boldsymbol \eta= \{\eta_{i,k}\}$ with $\eta_{i,k}$ defined analogously to \cref{eq:eta_k_App} with an additional index $i$ for each spin.
We can repeat the same steps as the previous subsection to integrate over the noise $\xi$, but this time we assume a general Wigner function ${\mathscr W}_0(\psi_0)$ and do not integrate over $\psi_0$. More explicitly, we separate out the dependence of the classical field $\underline \psi$ on noise and the initial conditions as $\underline \psi (t)= \psi[\xi](t)+ iG(t)\psi_0$ where $\psi[\xi](t)= \int_{t'}G(t-t')\xi(t')$ is solely the contribution due to noise. 
Now integrating over noise $\xi$, we obtain
\begin{align}
    \begin{split}
    \langle \vec \sigma| \rho_{S}(t)|\vec\sigma'\rangle =
    &\int d^2\psi_0 {\mathscr W}_0(\psi_0)
     \sum_{{\boldsymbol\eta},\boldsymbol{\tilde\eta}} \llangle \vec \sigma, \vec \sigma'|\prod_i \eta_{i,n}\tilde\eta_{i,n}\rrangle  \llangle \eta_{i,0}\tilde\eta_{i,0} |\rho_i(0)\rrangle \prod_{i,k}\llangle \eta_{i,k+1}\tilde\eta_{i,k+1}|e^{\delta t\mathbb L_{S}}|\eta_{i,k}\tilde\eta_{i,k} \rrangle  \\
    &\times e^{-i  \sum_i \frac{g_i}{\sqrt{N}}  \int_t \tilde\eta_{i}(t)(iG(t)\psi_0+c.c.)
     -\sum_{ij}\frac{g_ig_j}{2N}\int_{t,t'}\tilde \eta_j(t)\tilde \eta_j(t')  {i\Delta \cal G^K}(t,t')-i\sum_{ij} \frac{g_ig_j}{N}\int_{t,t'}\tilde \eta_i(t)\eta_j(t'){\cal G}(t,t')} \label{eq:Z_2nd_line}  \end{split}
\end{align}    
where we have used the continuous time as a shorthand for discrete time,
and defined 
\begin{equation}
   i \Delta  {\cal G}^K(t,t') = \overline{\psi[\xi](t) \bar{\psi}[\xi](t')} +c.c. =i{\cal G}^K(t,t') - [\frac{1}{2}G(t)\bar G(t') +c.c.]
\end{equation}
where the last equality follows by subtracting the contribution due to the initial state. 
We also remark that the all-to-all coupling (albeit with different coefficients) in \cref{eq:Z_2nd_line} is simply because the same bosonic mode is coupled to all the spins. 

Next, we provide an alternative representation that yields the same functional integral as \cref{eq:Z_2nd_line}, but it has the advantage that different spins become decoupled.
We first introduce  the independent noise variables $\bxi=\{\xi_{i,k}\}$:
\begin{equation}
    \overline{\xi_{i,k} \bar\xi_{j,k'}} =\frac{\kappa}{\delta t}\delta_{kk'}\delta_{ij}
\end{equation}
We then separate out the initial conditions as $\underline\psi (t)= \psi[\bxi](t) + iG(t) \psi_0$ where the term $\psi[\bxi]$, still to be defined, is linearly dependent on the new noise variables.
We are then tasked to define $\psi[\bxi]$ such that $(\overline{\psi[\bxi](t) \xi_i(t')}+c.c.)/2\kappa = {\cal G}(t-t')$ while $\overline{\psi[\bxi](t) \bar\psi[\bxi](t')}+c.c. = i\Delta {\cal G}^K(t,t')$; if possible, this would allow us to couple the spin $i$ only to $\xi_i$ and still recover \cref{eq:Z_2nd_line}. Indeed, this can be achieved by defining (in continuous time)
\begin{equation}
    {G}^{-1} \underline\psi(t) =\frac{1}{\sqrt{N}} \sum_j \xi_j(t)\qquad \mbox{or} \qquad \psi[\bxi](t) = \frac{1}{\sqrt{N}} \sum_j \int_{t'} G(t-t') \xi_j(t')
\end{equation}
where $G^{-1}=i\partial_t -\omega+i\kappa$ is the inverse of the Green's function $G$.
We can then construct the functional integral in discrete time as 
\begin{align}
    \begin{split}
            \langle \vec \sigma| \rho_{S}(t)|\vec\sigma'\rangle  = 
    &\int d^2\psi_0 {\mathscr W}_0(\psi_0)\int {\mathscr D}[\boldsymbol \xi] e^{-\sum_{i,k} \bar \xi_{i,k} \xi_{i,k}/\kappa}\sum_{\boldsymbol\eta,\boldsymbol{\tilde\eta}}\llangle \vec \sigma\vec\sigma'|\prod_i \eta_{i,n}\tilde\eta_{i,n}\rrangle  \llangle \eta_{i,0}\tilde\eta_{i,0} |\rho_i(0)\rrangle  \times \\
    &\times\prod_{i,k}\llangle \eta_{i,k+1}\tilde\eta_{i,k+1}|e^{\delta t\mathbb L_{S}}|\eta_{i,k}\tilde\eta_{i,k} \rrangle  e^{-i\delta t  \frac{g_i}{\sqrt{N}} \tilde\eta_{i,k} (\underline\psi_k+\underline{\bar\psi}_k)+\frac{g_i}{2\kappa}\delta t  \eta_{i,k}(\xi_{i,k}+\bar\xi_{i,k})-\frac{g_i^2}{4\kappa} \eta_{i,k}^2 } 
    \end{split}
\end{align}
where we have further assumed that the spins are initially factorized. One can explicitly check that the integral over noise $\xi_{i,k}$ exactly yields \cref{eq:Z_2nd_line}. 
But in this process, we have substituted the square of the sum
\(    
\big(\sum_j g_j\eta_j(t) \big)^2
\) with the sum of squares
\(
    \sum_j g_j^2\eta_j(t) ^2 
\)
where different spins are uncoupled. Indeed, undoing the insertion of the identity matrices, we obtain the stochastic quantum evolution with the dynamics generator given by \cref{eq:K_I_N_spin}. Therefore, we can evolve each spin independently before averaging their product over noise. 

\subsection{The case $N=1$ and $M>1$}\label{sec:FV_3}
Here, we consider a spin coupled to $M$ bosons. We  first consider a factorized state and assume that bosons are initially in their ground state; we shall later relax the latter assumption. 
Our starting point is \cref{eq:rho_s_App}  
upon substituting  $\xi \to \Xi$ and $\underline\psi \to \underline\Psi$ whose correlations are given by \cref{eq:chi_and_C_2}.
Following the same steps as \cref{sec:FV_1}, we find the expression in \cref{eq:Zeta} for the Feynman-Vernon influence functional, only with a different identification of the Green's functions as 
\begin{align}\label{eq:GR&GK}
    \begin{split}
    &{\cal G}(t,t')
    =  -\sum g_\alpha^2\sin[\omega_\alpha (t-t')]e^{-\kappa_\alpha(t-t')}\Theta(t-t') \\
    &i{\cal G}^K(t,t')= \sum g_\alpha^2 \cos[\omega_\alpha (t-t')] e^{-\kappa_\alpha|t-t'|} 
    \end{split}
\end{align}
In the absence of Markovian dissipation, the standard Feynman-Vernon influence functional is given by 
\begin{equation}\label{eq:FV_influence_func}
    F[\eta, \tilde\eta]= \exp \left\{-\frac{1}{\pi}\int_0^\infty dt \int^t_0 dt' \left[-iL_1(t-t') \eta(t) \tilde\eta(t')+L_2(t-t') \tilde \eta(t) \tilde \eta(t')\right]\right\}
\end{equation}
where the kernels $L_{1,2}(t,t')$ are defined in terms of the bath spectral function $J(\omega)$ as (at zero temperature)
\begin{align}
    \begin{split}
   L_1(t) = \int_0^\infty d\omega J(\omega) \sin(\omega t), \qquad 
   L_2(t) = \int_0^\infty d\omega J(\omega) \cos(\omega t)  
    \end{split}
\end{align}
Setting $\kappa\to0$, we recover the above expressions from \cref{eq:GR&GK} by identifying 
\begin{equation}
    \sum_\alpha \to \int \frac{d\omega}{\pi}, \qquad g_\alpha^2 \to J(\omega)   
\end{equation}

Next, we consider a general initial state for bosons. We derive \cref{eq:modified_modified_C} by combining the steps in this and the previous subsections. As a first step, we  define $\Psi[\Xi,\bX]$ from \cref{eq:modified_modified_C} by excluding the dependence on $\psi_{\alpha 0}$, that is, $\underline\Psi(t)= \Psi[\Xi, \bX](t) +i \sum_\alpha G_\alpha(t) \psi_{\alpha 0}$. 
With the correlations defined in \cref{eq:modified_modified_C}, one can show that  
    \begin{align}
    \begin{split}
     \overline{\Psi[\Xi,\bX](t) \bar \Psi[\Xi,\bX](t')} 
     &= \sum_\alpha g_\alpha^2 \kappa_\alpha \int_{t''} G_\alpha(t,t'') \bar G_\alpha(t',t'') =\sum_\alpha g_\alpha^2 \overline{\psi_\alpha[\xi_\alpha](t) \bar \psi_\alpha[\xi_\alpha](t')} \\
     \overline{\Psi[\Xi,\bX](t) \bar \Xi(t')}  &= \sum_\alpha \frac{g_\alpha^2}{2} G_\alpha(t-t') = \sum_\alpha g_\alpha^2 \overline{\psi_\alpha[\xi_\alpha](t) \bar \xi_\alpha(t')} 
     \end{split}
    \end{align} 
where $\psi_\alpha[\xi_\alpha](t)= \int_{t'} G_\alpha(t-t') \xi_\alpha(t')$. One can then follow the same steps as in \cref{sec:FV_2} to show that explicitly integrating over the noise $\Xi$ yields the same result as the  original functional integral over $M$ different noise variables $\{\xi_\alpha(t)\}$. In both cases, the initial state is represented by the integral over $\{\psi_{\alpha 0}\}$ weighted by the Wigner function ${\mathscr W}_0(\{\psi_{\alpha0}\})$.

\section{$\check C$ matrix}\label{sec:checkC}
In this section, we first prove that $\check C(t,t')$ defined in \cref{eq:check_C}, viewed as a matrix in time, is positive. We then provide a simple scheme for diagonalizing this matrix. 

To show the positivity, we first write $\check C$ as 
\begin{equation}\label{eq:cehckC_App}
    \check C(t,t') =  C_{\rm st}(t-t') + D(t,t')
\end{equation}
where we have defined
\begin{align}
     \begin{split}
     C_{\rm st}(t-t') &= C(t-t') - \frac{1}{\gamma}\int_{-\infty}^\infty ds  \chi(t-t'-s) \bar\chi(s)  \\
    D(t,t') &= \frac{1}{\gamma} \int_{-\infty}^0 dt''\chi(t-t'') \bar \chi(t'-t'') 
    \end{split}
\end{align}
Notice the bounds of the integrals in these expressions. The subscript st stands for the  stationary state since $\check C(t,t') \to C_{\rm st}(t,t')$ at long times $t,t'\to \infty$.
It is easy to see that $ D$ is manifestly positive:  $\langle f |  D|f\rangle \equiv \int_{t,t'} \bar f(t) D(t,t')f(t') =(1/\gamma)\int dt'' \left|\int_t f(t) \chi(t-t'')\right|^2\ge 0$ for an arbitrary function $f(t)$. To prove $\check C >0$, it is sufficient to show that $C_{\rm st} \ge 0$. Since the latter is time translation invariant, we can consider its Fourier transform
\begin{equation}
    C_{\rm st}(\omega) =\sum_\alpha g_\alpha^2 \kappa_\alpha {G_\alpha(\omega)} \overline{G_\alpha(\omega)}   - \frac{1}{\sum g_\alpha^2/\kappa_\alpha}\sum_\alpha g_\alpha^2 {G_\alpha (\omega)}  \sum_\beta g_\beta^2 \overline{G_\beta (\omega)} 
\end{equation}
This quantity is positive because
\begin{align}
\begin{split}
    \gamma  C_{\rm st}(\omega) 
    &= \frac{1}{2}\sum_{\alpha,\beta} g_\alpha^2 g_\beta^2 \left[\frac{\kappa_\alpha}{\kappa_\beta} |G_\alpha(\omega)|^2 + \frac{\kappa_\beta}{\kappa_\alpha} |G_\beta(\omega)|^2 - G_\alpha (\omega)\overline{ G_\beta (\omega)}- \overline{ G_\alpha(\omega)}  G_\beta(\omega)   \right] \\
    &= \frac{1}{2} \sum_{\alpha,\beta} g_\alpha^2 g_\beta^2 \left|\sqrt{\frac{\kappa_\alpha}{\kappa_\beta}} G_\alpha(\omega)-\sqrt{\frac{\kappa_\beta}{\kappa_\alpha}} G_\beta(\omega)\right|^2 \ge 0 \label{eq:C_positive}
\end{split}
\end{align}
 
We therefore conclude that $\check C(t,t')$ is positive as a matrix. 

In fact, a similar argument leads to the stronger result that  
$\Delta C= \check C(t,t') - \frac{1}{2}\sum g_\alpha^2 G_\alpha(t) \bar G_\alpha(t')$ is positive, that is, 
\begin{align}
     \Delta  C(t,t')=\sum_\alpha g^2_\alpha \kappa_\alpha \int_{t''} G_\alpha(t-t'') \bar G_\alpha(t'-t'') -\frac{1}{\gamma} \int_{t''} \chi(t-t'') \bar \chi(t'-t'') \ge 0
\end{align}
in the sense of a matrix. The proof follows by sandwiching the above equation between $\langle f|$ and $|f\rangle$ for an arbitrary function $f(t)$ and following a similar argument as above. Note that $\Delta C$ appears in \cref{eq:modified_modified_C} as the colored noise correlator upon excluding the contribution from the initial state.

Next, we diagonalize $\check C(t,t')$ viewed as a matrix in the time domain. It is more convenient to work in the Fourier basis. 
We first Fourier transform the function $C_{\rm st}(t-t')$. Defining the maximum time $t_{max}$, the argument of $C_{\rm st}$ is in the range $[-t_{max}, t_{max}]$; see also \cite{Peter_Stochastic_Schrodinger_Equation,kamar_spin_boson_model_stochastic_SCH}. Expanding in Fourier series, we have  
\begin{equation}
     C_{\rm st}(t-t') =\sum_{n=-\infty}^\infty c_n e^{i\pi n (t-t')/t_{max}}
\end{equation}
where the coefficients $c_n$ are obtained as
\begin{equation}
    c_n = \frac{1}{2t_{max}}\int_{-t_{max}}^{t_{max}} ds e^{-i\pi n s/t_{max}}  C_{\rm st}(s) 
\end{equation}
Next, we consider the function $D(t,t')$; the time variables $t,t'$ are defined in the range $[0,t_{max}]$. However, we extend these functions to the extended domain $[-t_{max},t_{max}]$, in order to expand these functions in the same harmonics basis used for $C_{\rm st}(t-t')$. There are different ways that one can define these functions in an extended domains. A first choice is to define $\check C(t,t')=0$ if $t<0$ or $t'<0$, although it may generate pronounced oscillations when we truncate the sum over harmonics at a finite order $n_{max}$. Other choices may be taken where the function $\check C(t,t')$ are nonzero for $t<0$ or $t'<0$. 
We expand $D(t,t')$ in terms of these harmonics as
\begin{equation}
     D(t,t')= \sum_{n,n' = -\infty}^\infty d_{nn'}  e^{i\pi nt/t_{max} -i\pi n't'/t_{max} } 
\end{equation}
where 
\begin{equation}
     d_{nn'} = \int\!\!\!\int_{-t_{max}}^{t_{max}} \frac{ds ds'}{(2t_{max})^2}   e^{-i\pi( n s-n' s')/t_{max}} D(s,s') 
\end{equation}
The full matrix can be then written as 
\begin{equation}
    \check C(t,t') = \sum_{n,n'} e^{i\pi n t /t_{max}-i\pi n' t'/t_{max}}  (c_n \delta_{n,n'} + d_{n,n'})
\end{equation}
We can now diagonalize the matrix $[\check {\mathbf C}]_{n,n'} = c_n \delta_{n,n'}+  d_{nn'}$ as
\begin{equation}
    \check {\mathbf C} = {\mathbf U}^{-1} \check {\mathbf C}_{\rm diag} {\mathbf U}
\end{equation}
where $\check {\mathbf C}_{\rm diag} = {\rm diag}\{\check c_a\}$ is a diagonal matrix with all the eigenvalues positive ($\check c_a\ge 0$), and $\mathbf U$ is a unitary matrix. 
Finally, we can write 
\begin{equation}
    \check C(t,t') = \sum_{a}{\check c}_a\bar\theta_a(t) \theta_a(t') 
\end{equation}
where the new functions $\theta_\alpha$ are now defined as
\begin{equation}
    \theta_a(t) = \sum_n U_{a n} e^{-i\pi n t/t_{max}} 
\end{equation}

\end{widetext}

% \bibliography{refs}
% \bibliographystyle{apsrev4-2}


\end{document}